\documentclass{article}
\usepackage{amsfonts}
\usepackage{amsmath}
\usepackage{graphicx}
\newtheorem{theorem}{Theorem}

\newtheorem{lemma}[theorem]{Lemma}

\newtheorem{proposition}[theorem]{Proposition}

\input{tcilatex}

\begin{document}

\begin{center}
{\Large \textbf{ The conformal geometry }}

{\Large \textbf{ of }}

{\Large \textbf{Random Regge Triangulations}}

\bigskip

{\large \textsl{M. Carfora}$\,^{b,}$\footnote{{\large email
mauro.carfora@pv.infn.it}},} \vspace{24pt}

\textsl{C}{\large \textsl{. Dappiaggi}$\,^{b,}$\footnote{{\large email
claudio.dappiaggi@pv.infn.it}},} \vspace{24pt}

{\large \textsl{A. Marzuoli}$\,^{b,}$\footnote{{\large email
annalisa.marzuoli@pv.infn.it}},} \vspace{24pt}

$^{b}$~Dipartimento di Fisica Nucleare e Teorica,

Universit\`{a} degli Studi di Pavia, \\[0pt]
via A. Bassi 6, I-27100 Pavia, Italy, \\[0pt]
and\\[0pt]
Istituto Nazionale di Fisica Nucleare, Sezione di Pavia, \\[0pt]
via A. Bassi 6, I-27100 Pavia, Italy
\end{center}

\newpage

\section{Introduction}

We present here a report on our efforts to understand the conformal geometry
associated with the use of simplicial methods in 2-dimensional quantum
gravity. In particular, we trace a logical path among the main characters
of such a theory: dynamical triangulations, Regge surfaces, and Riemann
moduli theory. The choice of such a subject seems appropriate in a volume of
papers dedicated to A. Lichnerowicz, since the interplay between conformal
geometry and gravity has always been an important motivation in his
research. Such an interplay  is crucial in 2-dimensional (quantum) gravity,
and it is safe to say that the successful analysis of such a theory can be
traced back to the special role that conformal geometry plays in dimension
two. However, when we enter a region of strong coupling regime between
matter and gravity, we do not yet have a complete understanding of the
field-theoretic dynamics of the conformal mode of the theory. This is an
important problem whose relevance goes far beyond the study of
two-dimensional gravity and string theories and has received considerable
attention in recent years. While we are still far from a satisfactory
solution, it is fair to observe that some clue to its understanding are
coming out from the use of techniques of Piecewise-Linear (PL) geometry. As
a matter of fact, triangulated surfaces provide one of the most powerful
techniques for analyzing two-dimensional quantum gravity in regimes which are
not accessible to the standard field-theoretic formalism. In such a sense,
it is important to establish a connection between conformal geometry and
quantum gravity which is more directly related with the use of simplicial
methods, thus making their role in the theory more explicit.

In this paper, which is a natural evolution of one of our previous works \cite{carfora}, 
we discuss such a topic by using surfaces endowed with
triangulations with variable connectivity and fluctuating egde-lengths. Such
piecewise-linear surfaces are not proper Regge triangulations, since their
adjacency matrix is not a priori fixed, nor they are dynamical
triangulations, since they are generated by glueing triangles which are not,
in general, equilateral. Random Regge Triangulations seems an appropriate
name \cite{carfora}. They allow for a relatively
simple and direct analyisis of the modular properties of 2 dimensional
simplicial quantum gravity, since such triangulations are naturally related
to the Weil-Petersson geometry of the (compactified) moduli space of genus $g$ 
Riemann surfaces with $N_{0}$ punctures $\overline{\mathfrak{M}}
_{g},_{N_{0}}$, (the number of punctures $N_{0}$ is the number of vertices
of the triangulations).  The main result of our analysis is the explicit
association of a Weil-Petersson metric to a Regge triangulation. With such a
metric at our disposal, we can formally evaluate the Weil-Petersson volume
over the space of all random Regge triangulations. By exploiting a recent
result of Manin and Zograf \cite{manin}, we can show that such a volume
provides the (large $N_{0}$ asymptotics of the) dynamical triangulation
partition function for pure gravity. We conclude the paper by discussing 
the (regularized) Liouville action associated with random Regge
triangulations and its connection with Hodge-Deligne theory. 

\noindent From a field theoretic point of view there is a tendency 
to relegate simplicial methods to the ancillary role of
a regularization scheme, playing a role in gravity similar to that of
lattice regularizations in gauge theories. However, according to some of the
results discussed in this paper, one cannot help thinking that in gravity their role is more
foundational, and that they rely on a set of first principles probably
connected with the holographic hypothesis \cite{arcioni},\cite{arcioni2}
\section{Triangulated surfaces and Ribbon graphs}

Let $T$ denote a $2$-dimensional simplicial complex with underlying
polyhedron $|T|$ and $f$- vector $(N_{0}(T),N_{1}(T),N_{2}(T))$, where $
N_{i}(T)\in \mathbb{N}$ is the number of $i$-dimensional sub- simplices $
\sigma ^{i}$ of $T$. Given a simplex $\sigma $ we denote by $st(\sigma )$,
(the star of $\sigma $), the union of all simplices of which $\sigma $ is a
face, and by $lk(\sigma )$, (the link of $\sigma $), is the union of all
faces $\sigma ^{f}$ of the simplices in $st(\sigma )$ such that $\sigma
^{f}\cap \sigma =\emptyset $. A random Regge triangulation of a $2$
-dimensional PL manifold $M$, (without boundary), is a homeomorphism $
|T_{l}|\rightarrow {M}\ $ where each face of $T$ is geometrically
realized by a rectilinear simplex of variable edge-lengths $l(\sigma ^{1}(k))
$. A dynamical triangulation $|T_{l=a}|\rightarrow {M}$ is a particular
case of a random Regge PL-manifold realized by rectilinear and equilateral
simplices of edge-length $l(\sigma ^{1}(k))=$ $a$. 

\begin{figure}[h]
\begin{center}
\includegraphics[scale=.4]{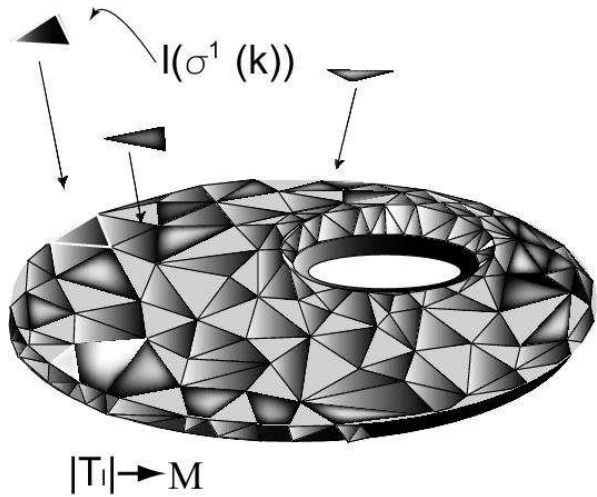}
\caption{A torus triangulated with triangles of variable edge-length.}
\end{center}
\end{figure}

The metric structure of a Regge triangulation is locally Euclidean everywhere except at the vertices $
\sigma ^{0}$, (the \textit{bones}), where the sum of the dihedral angles, $
\theta (\sigma ^{2})$, of the incident triangles $\sigma ^{2}$'s is in
excess (negative curvature) or in defect (positive curvature) with respect
to the $2\pi $ flatness constraint. The corresponding deficit angle $
\varepsilon $ is defined by $\varepsilon =2\pi -\sum_{\sigma ^{2}}\theta
(\sigma ^{2})$, where the summation is extended to all $2$-dimensional
simplices incident on the given bone $\sigma ^{0}$. If $K_{T}^{0}$ denotes
the $(0)$-skeleton of $|T_{l}|\rightarrow {M}$, (\emph{i.e.}, the collection
of vertices of the triangulation), then $M\backslash {K_{T}^{0}}$ is a
flat Riemannian manifold, and any point in the interior of an $r$- simplex $
\sigma ^{r}$ has a neighborhood homeomorphic to $B^{r}\times {C}(lk(\sigma
^{r}))$, where $B^{r}$ denotes the ball in $\mathbb{R}^{n}$ and ${C}
(lk(\sigma ^{r}))$ is the cone over the link $lk(\sigma ^{r})$, (the product 
$lk(\sigma ^{r})\times \lbrack 0,1]$ with $lk(\sigma ^{r})\times \{1\}$
identified to a point). In particular, let us denote by $C|lk(\sigma
^{0}(k))|$ the cone over the link of the vertex $\sigma ^{0}(k)$. On any
such a disk $C|lk(\sigma ^{0}(k))|$ we can introduce a locally uniformizing
complex coordinate $\zeta (k)\in \mathbb{C}$ in terms of which we can
explicitly write down a conformal conical metric locally characterizing the
singular structure of $|T_{l}|\rightarrow M$, \emph{viz.}, 
\begin{equation}
e^{2u}\left| \zeta (k)-\zeta _{k}(\sigma ^{0}(k))\right| ^{-2\left( \frac{
\varepsilon (k)}{2\pi }\right) }\left| d\zeta (k)\right| ^{2},  \label{cmetr}
\end{equation}
where $\varepsilon (k)$ is the corresponding deficit angle, and $
u:B^{2}\rightarrow \mathbb{R}$ is a continuous function ($C^{2}$ on $
B^{2}-\{\sigma ^{0}(k)\}$) such that, for $\zeta (k)\rightarrow \zeta
_{k}(\sigma ^{0}(k))$, we have $\left| \zeta (k)-\zeta _{k}(\sigma
^{0}(k))\right| \frac{\partial u}{\partial \zeta (k)}$, and $\left| \zeta
(k)-\zeta _{k}(\sigma ^{0}(k))\right| \frac{\partial u}{\partial \overline{
\zeta }(k)}$ both $\rightarrow 0$. Up to the presence of $e^{2u}$, we
immediately recognize in such an expression the metric $g_{\theta (k)}$ of a
Euclidean cone of total angle $\theta (k)=2\pi -\varepsilon (k)$. The factor 
$e^{2u}$ allows to move within the conformal class of all metrics possessing
the same singular structure of the triangulated surface $|T_{l}|\rightarrow M$.

\begin{figure}[ht] 
\begin{center}
\includegraphics[scale=.4]{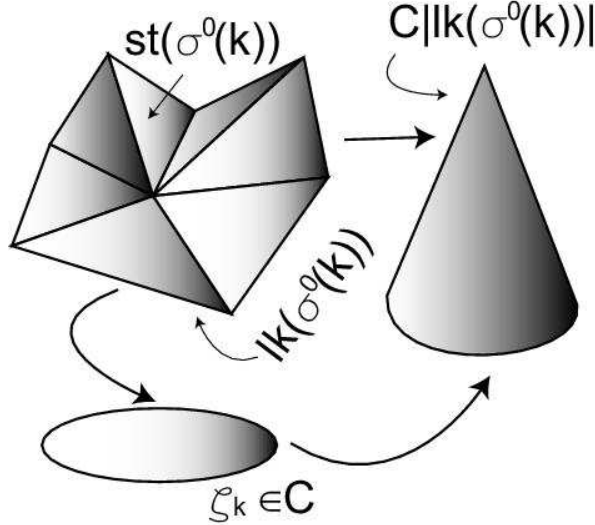}
\caption{The geometric structures around a vertex.}
\end{center}
\end{figure}

\noindent We shall denote by 
\begin{equation}
\left[ g_{\theta (k)}\right] \doteq \left\{ e^{2u}g_{\theta (k)}\,:\,u\in
C^{0}(B^{2})\cap C^{2}(B^{2}-\{\sigma ^{0}(k)\})\right\} 
\end{equation}
the conformal class determined by the conical metric $g_{\theta (k)}$. If we
allow for conformal factors with logarithmic singularities then we can
equivalently consider the conformal class of the metric representing the
Regge triangulation as given by $ds^{2}=e^{2v}ds_{0}^{2}$ where $ds_{0}^{2}$
is a smooth metric on $M$ and where the conformal factor $v$ around the
generic vertex $\sigma ^{0}(k)$, is provided by 
\begin{equation}
\left. v\right| _{U_{\rho ^{2}(k)}}=-\frac{\varepsilon (k)}{2\pi }\ln \left|
\zeta (k)-\zeta _{k}(\sigma ^{0}(k))\right| +u.
\end{equation}
We can profitably shift between these two points of view by exploiting
standard techniques of complex analysis, and making contact with moduli
space theory.

\bigskip

\subsection*{\normalsize{\bf 2.1 Curvature assignments and divisors. }} In the case of
dynamical triangulations, the picture simplifies considerably since the
deficit angles are generated by the numbers $\#\{\sigma ^{2}(h)\bot \sigma
^{0}(i)\}$ of triangles incident on the $N_{0}(T)$ vertices, the \textit{
curvature assignments}, $\{q(k)\}_{k=1}^{N_{0}(T)}\in \mathbb{N}^{N_{0}(T)}$,

\begin{equation}
q(i)=\frac{2\pi -\varepsilon (i)}{\arccos (1/2)}.  \label{curvat}
\end{equation}
For a regular triangulation we have $q(k)\geq 3$, and since each triangle
has $3$ vertices $\sigma ^{0}$, the set of integers $\{q(k)
\}_{k=1}^{N_{0}(T)}$ is constrained by

\begin{equation}
\sum_{k}^{N_{0}}q(k)=3N_{2}=6\left[ 1-\frac{\chi (M)}{N_{0}(T)}\right]
N_{0}(T),  \label{vincolo}
\end{equation}
where $\chi (M)$ denotes the Euler-Poincar\'{e} characteristic of the
surface, and where $6\left[ 1-\frac{\chi (M)}{N_{0}(T)}\right] $, ($\simeq 6$
for $N_{0}(T)>>1$), is the average value of the curvature assignments $\{q(k)\}_{k=1}^{N_{0}}$. 
More generally we shall consider semi-simplicial
complexes for which the constraint $q(k)\geq 3$ is removed. Examples of such
configurations are afforded by triangulations with pockets, where two
triangles are incident on a vertex, or by triangulations where the star of a
vertex may contain just one triangle. We shall refer to such extended
configurations as generalized (Regge and dynamical) triangulations.

The singular structure of the metric defined by (\ref{cmetr}) can be
naturally summarized in a formal linear combination of the points $\{\sigma
^{0}(k)\}$ with coefficients given by the corresponding deficit angles
(normalized to $2\pi $), \emph{viz.}, in the \emph{real divisor }\cite{troyanov} 
\begin{equation}
Div(T)\doteq \sum_{k=1}^{N_{0}(T)}\left( -\frac{\varepsilon (k)}{2\pi }
\right) \sigma ^{0}(k)=\sum_{k=1}^{N_{0}(T)}\left( \frac{\theta (k)}{2\pi }
-1\right) \sigma ^{0}(k)
\end{equation}
supported on the set of bones $\{\sigma ^{0}(i)\}_{i=1}^{N_{0}(T)}$. Note
that the degree of such a divisor, defined by 
\begin{equation}
\left| Div(T)\right| \doteq \sum_{k=1}^{N_{0}(T)}\left( \frac{\theta (k)}{
2\pi }-1\right) =-\chi (M)  \label{rediv}
\end{equation}
is, for dynamical triangulations, a rewriting of the combinatorial
constraint (\ref{vincolo}). In such a sense, the pair $(|T_{l=a}|\rightarrow
M,Div(T))$, or shortly, $(T,Div(T))$, encodes the datum of the triangulation 
$|T_{l=a}|\rightarrow M$ and of a corresponding set of curvature assignments 
$\{q(k)\}$ on the vertices $\{\sigma ^{0}(i)\}_{i=1}^{N_{0}(T)}$. The real
divisor $\left| Div(T)\right| $ characterizes the Euler class of the pair 
$(T,Div(T))$ and yields for a corresponding Gauss-Bonnet formula.
Explicitly, the Euler number associated with $(T,Div(T))$ is defined, \cite{troyanov},
by

\begin{equation}
e(T,Div(T))\doteq \chi (M)+|Div(T)\mathbf{|.}  \label{euler}
\end{equation}
and the Gauss-Bonnet formula reads \cite{troyanov}:

\begin{lemma}
(\textbf{Gauss-Bonnet for triangulated surfaces}) Let $(T,Div(T))$ be a
triangulated surface with divisor 
\begin{equation}
Div(T)\doteq \sum_{k=1}^{N_{0}(T)}\left( \frac{\theta (k)}{2\pi }-1\right)
\sigma ^{0}(k),
\end{equation}
associated with the vertices $\{\sigma ^{0}(k)\}_{k=1}^{N_{0}(T)}$. Let $
ds^{2}$ be the conformal metric (\ref{cmetr}) representing the divisor $
Div(T)$ . Then 
\begin{equation}
\frac{1}{2\pi }\int_{M}KdA=e(T,Div(T)),  \label{Euclass}
\end{equation}
where $K$ and $dA$ respectively are the curvature and the area element
corresponding to the local metric $ds_{(k)}^{2}.$
\end{lemma}

Note that such a theorem holds for any singular Riemann surface $\Sigma $
described by a divisor $Div(\Sigma )$ and not just for triangulated surfaces
\cite{troyanov}. Since for a Regge (dynamical) triangulation, we have $
e(T_{a},Div(T))=0 $, the Gauss-Bonnet formula implies

\begin{equation}
\frac{1}{2\pi }\int_{M}KdA=0.  \label{GaussB}
\end{equation}
Thus, a triangulation $|T_{l}|\rightarrow M$ naturally carries a conformally
flat structure. Clearly this is a rather obvious result, (since the metric
in $M-\{\sigma ^{0}(i)\}_{i=1}^{N_{0}(T)}$ is flat). However, it admits a
not-trivial converse (recently proved by M. Troyanov, but, in a sense, going
back to E. Picard) \cite{troyanov}, \cite{picard}:

\begin{theorem}
(\textbf{Troyanov-Picard}) Let $(\left( M,\mathcal{C}_{sg}\right) ,Div)$ be
a singular Riemann surface with a divisor such that $e(M,Div)=0$. Then there
exists on $M$ a unique (up to homothety) conformally flat metric
representing the divisor $Div$.
\end{theorem}

\subsection*{\normalsize{\bf 2.2. Conical Regge polytopes.}}Let us consider the (first)
barycentric subdivision of $T$. The closed stars, in such a subdivision,
of the vertices of the original triangulation $T_{l}$ form a collection of $
2 $-cells $\{\rho ^{2}(i)\}_{i=1}^{N_{0}(T)}$ characterizing the \emph{
conical} Regge polytope $|P_{T_{l}}|\rightarrow {M}$ (and its rigid
equilateral specialization $|P_{T_{a}}|\rightarrow {M}$) barycentrically
dual to $|T_{l}|\rightarrow {M}$. If $(\lambda (k),\chi (k))$ denote polar
coordinates (based at $\sigma ^{0}(k)$) of a point $p\in \rho ^{2}(k)$, then 
$\rho ^{2}(k)$ is geometrically realized as the space 
\begin{equation}
\left. \left\{ (\lambda (k),\chi (k))\ :\lambda (k)\geq 0;\chi (k)\in 
\mathbb{R}/(2\pi -\varepsilon (k))\mathbb{Z}\right\} \right/ (0,\chi
(k))\sim (0,\chi ^{\prime }(k))
\end{equation}
endowed with the metric 
\begin{equation}
d\lambda (k)^{2}+\lambda (k)^{2}d\chi (k)^{2}.
\end{equation}

\begin{figure}[ht]
\begin{center}
\includegraphics[scale=.4]{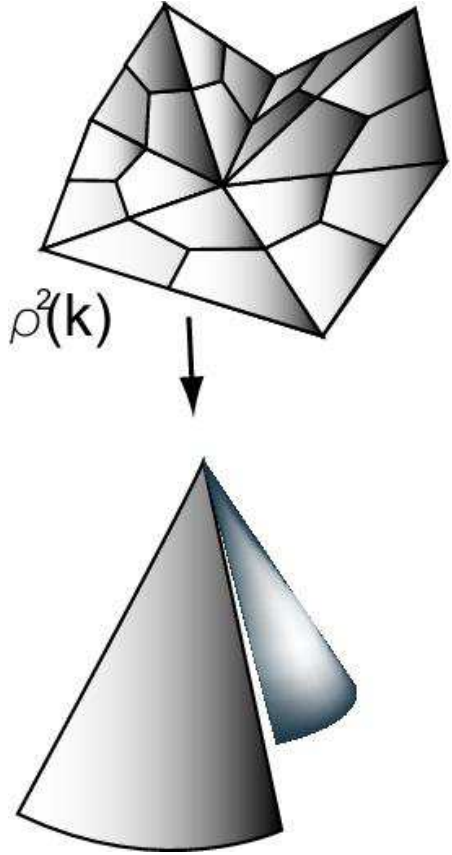}
\caption{The conical geometry of the baricentrically dual polytope.}
\end{center}
\end{figure}

\noindent In other words, here we are not considering a rectilinear presentation of
the dual cell complex $P$ (where the PL-polytope is realized by flat
polygonal $2$-cells $\{\rho ^{2}(i)\}_{i=1}^{N_{0}(T)}$) but rather a
geometrical presentation $|P_{T_{l}}|\rightarrow {M}$ of $P$ where the $2$
-cells $\{\rho ^{2}(i)\}_{i=1}^{N_{0}(T)}$ retain the conical geometry
induced on the barycentric subdivision by the original metric (\ref{cmetr})
structure of $|T_{l}|\rightarrow {M}$.

\bigskip

\subsection*{\normalsize{\bf 2.3. Hyperbolic cusps and cylindrical ends}}
It is important to stress that wheras a Regge triangulation characterizes
a unique (up to automorphisms) singular Euclidean structure, this latter
actually allows for a more general type of metric triangulation. The point
is that some of the vertices associated with a singular Euclidean structure
can be characterized by deficit angles $\varepsilon (k)\rightarrow 2\pi $,
\emph{\ i.e.}, $\sum_{\sigma ^{2}(k)}\theta (\sigma ^{2}(k))=0$, where the
summation is extended to all triangles incident on the given vertex $\sigma
^{0}(k)\in |T_{l}|\rightarrow {M}$. Such a situation corresponds to having
the cone $C|lk(\sigma ^{0}(k))|$ over the link $lk(\sigma ^{0}(k))$
realized by a Euclidean cone of angle $0$. This is a natural limiting case
in a Regge triangulation, (think of a vertex where many long and thin
triangles are incident), and it is usually discarded as an unwanted
pathology. However, there is really nothing pathological about that. It can
be easily handled, since the corresponding $2$-cell $\rho ^{2}(k)\in
|P_{T_{l}}|\rightarrow {M}$ can be naturally endowed with the geometry of
a hyperbolic cusp, \emph{i.e.}, that of a half-infinite cylinder $\mathbb{S}
^{1}\times \mathbb{R}^{+}$ equipped with the hyperbolic metric $\lambda
(k)^{-2}(d\lambda (k)^{2}+d\chi (k)^{2})$. The triangles incident on $\sigma
^{0}(k)$ are then realized as hyperbolic triangles with the vertex $\sigma
^{0}(k)$ located at $\lambda (k)=\infty $ and corresponding angle $\theta
_{k}=0$ \cite{judge}. Alternatively, and perhaps more in the
spirit of Regge calculus, one may consider $\rho ^{2}(k)$ endowed with the
conformal Euclidean structure obtained from (\ref{cmetr}) by setting $\frac{
\varepsilon (k)}{2\pi }=1$, \emph{i.e.} 
\begin{equation}
e^{2u}\left| \zeta (k)-\zeta _{k}(\sigma ^{0}(k))\right| ^{-2}\left| d\zeta
(k)\right| ^{2},
\end{equation}
which (up to the conformal factor $e^{2u}$) is the flat metric on the
half-infinite cylinder $\mathbb{S}^{1}\times \mathbb{R}^{+}$ (a cylindrical
end). 

\begin{figure}[ht]
\begin{center}
\includegraphics[scale=.4]{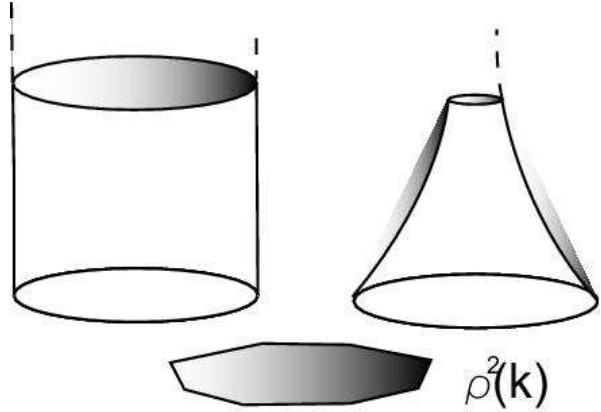}
\caption{The cylindrical and hyperbolic metric over a $\theta\to 0$ degenerating polytopal cell.}
\end{center}
\end{figure}

\noindent Since the Poincar\'{e} metric on the punctured disk 
$$D^2(k)=\{\zeta (k)\in
C|\;0<\left| \zeta (k)-\zeta _{k}(\sigma ^{0}(k))\right| <1\}$$
 is 
\begin{equation}
\left( \left| \zeta (k)-\zeta _{k}(\sigma ^{0}(k))\right| \ln \frac{1}{
\left| \zeta (k)-\zeta _{k}(\sigma ^{0}(k))\right| }\right) ^{-2}\left|
d\zeta (k)\right| ^{2},
\end{equation}
one can shift from the Euclidean to the hyperbolic metric by setting 
\begin{equation}
e^{2u}=\left( \ln \frac{1}{\left| \zeta (k)-\zeta _{k}(\sigma
^{0}(k))\right| }\right) ^{-2},
\end{equation}
and the two points of view are fully equivalent. 
At any rate the presence of such defects (hyperbolic cusps or cylindrical ends) is consistent with a
singular Euclidean structure as long as the associated divisor satisfies the
topological constraint (\ref{rediv}), which we can rewrite as 
\begin{equation}
\sum_{\{\frac{\varepsilon (k)}{2\pi }\neq 1\}}\left( -\frac{\varepsilon (k)}{
2\pi }\right) =2g-2+\#\left\{ \sigma ^{0}(h)|\;\frac{\varepsilon (h)}{2\pi }
=1\right\} .
\end{equation}
In particular, we can have the limiting case of the singular Euclidean
structure associated with a genus $g$ surface triangulated with $N_{0}-1$
hyperbolic vertices $\{\sigma ^{0}(k)\}_{k=1}^{N_{0}-1}$ (or, equivalently,
with $N_{0}-1$ cylindrical ends)\ and just one standard conical vertex, $
\sigma ^{0}(N_{0})$, supporting the deficit angle 
\begin{equation}
-\frac{\varepsilon (N_{0})}{2\pi }=2g-2+(N_{0}-1).  \label{lastmarked}
\end{equation}

\subsection*{\normalsize{\bf 2.4. Ribbon graphs.}} The geometrical realization of the $1$-skeleton 
of the conical Regge polytope $|P_{T_{l}}|\rightarrow {M}$ is a 
$3$-valent graph 
\begin{equation}
\Gamma =(\{\rho ^{0}(k)\},\{\rho ^{1}(j)\})
\end{equation}
where the vertex set $\{\rho ^{0}(k)\}_{k=1}^{N_{2}(T)}$ is identified with
the barycenters of the triangles $\{\sigma ^{o}(k)\}_{k=1}^{N_{2}(T)}\in
|T_{l}|\rightarrow M$, whereas each edge $\rho ^{1}(j)\in \{\rho
^{1}(j)\}_{j=1}^{N_{1}(T)}$ is generated by two half-edges $\rho ^{1}(j)^{+}$
and $\rho ^{1}(j)^{-}$ joined through the barycenters $\{W(h)
\}_{h=1}^{N_{1}(T)}$ of the edges $\{\sigma ^{1}(h)\}$ belonging to the
original triangulation $|T_{l}|\rightarrow M$. If we formally introduce a
ghost-vertex of a degree $2$ at each middle point $\{W(h)\}_{h=1}^{N_{1}(T)}$
, then the actual graph naturally associated to the $1$-skeleton of 
$|P_{T_{l}}|\rightarrow {M}$ is the edge-refinement \cite{mulase} of $\Gamma =(\{\rho
^{0}(k)\},\{\rho ^{1}(j)\})$, \emph{i.e.} 
\begin{equation}
\Gamma _{ref}=\left( \{\rho
^{0}(k)\}\bigsqcup_{h=1}^{N_{1}(T)}\{W(h)\},\{\rho
^{1}(j)^{+}\}\bigsqcup_{j=1}^{N_{1}(T)}\{\rho ^{1}(j)^{-}\}\right) .
\end{equation}
The natural automorphism group $Aut(P_{l})$ of $|P_{T_{l}}|\rightarrow {M}$
, (\emph{i.e.}, the set of bijective maps $\Gamma =(\{\rho ^{0}(k)\},\{\rho
^{1}(j)\})\rightarrow \widetilde{\Gamma }=(\widetilde{\{\rho ^{0}(k)\}},
\widetilde{\{\rho ^{1}(j)\}}$ preserving the incidence relations defining
the graph structure), is the automorphism group of its edge refinement \cite{mulase}, 
$Aut(P_{l})\doteq Aut(\Gamma _{ref})$. The locally uniformizing complex
coordinate $\zeta (k)\in \mathbb{C}$ in terms of which we can explicitly
write down the singular Euclidean metric (\ref{cmetr}) around each vertex $
\sigma ^{0}(k)\in $ $|T_{l}|\rightarrow M$, provides a (counterclockwise)
orientation in the $2$-cells of $|P_{T_{l}}|\rightarrow {M}$. Such an
orientation gives rise to a cyclic ordering on the set of half-edges $\{\rho
^{1}(j)^{\pm }\}_{j=1}^{N_{1}(T)}$ incident on the vertices $\{\rho
^{0}(k)\}_{k=1}^{N_{2}(T)}$. 

\begin{figure}[ht]
\begin{center}
\includegraphics[scale=.4]{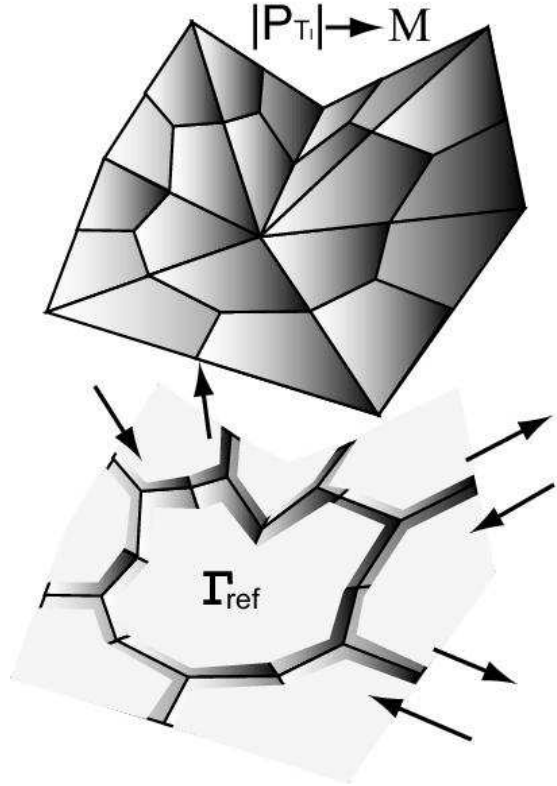}
\caption{The dual polytope around a vertex and it's edge refinement.}
\end{center}
\end{figure}

According to these remarks, the $1$-skeleton
of $|P_{T_{l}}|\rightarrow {M}$ is a ribbon (or fat) graph \cite{ambjorn}, \emph{viz.}, 
a graph $\Gamma $ together with a cyclic ordering on the set of half-edges
incident to each vertex of $\Gamma$. Conversely, any ribbon graph $\Gamma 
$ characterizes an oriented surface $M(\Gamma )$ with boundary possessing $
\Gamma $ as a spine, (\emph{i.e.}, the inclusion $\Gamma \hookrightarrow
M(\Gamma )$ is a homotopy equivalence). In this way (the edge-refinement of)
the $1$-skeleton of a generalized conical Regge polytope 
$|P_{T_{l}}|\rightarrow {M}$ is in a one-to-one correspondence with trivalent
metric ribbon graphs. The set of all such trivalent ribbon graph $\Gamma $
with given edge-set $e(\Gamma )$\ can be characterized \cite{mulase}, \cite{loojenga} as a space
homeomorphic to $\mathbb{R}_{+}^{|e(\Gamma )|}$, ($|e(\Gamma )|$ denoting
the number of edges in $e(\Gamma )$), topologized by the standard 
$\epsilon $-neighborhoods $U_{\epsilon }\subset $ $\mathbb{R}
_{+}^{|e(\Gamma )|}$. The automorphism group $Aut(\Gamma )$ acts naturally
on such a space via the homomorphism $Aut(\Gamma )\rightarrow \mathfrak{G}
_{e(\Gamma )}$, where $\mathfrak{G}_{e(\Gamma )}$ denotes the symmetric group
over $|e(\Gamma )|$ elements, and the resulting quotient space $\mathbb{R}
_{+}^{|e(\Gamma )|}/Aut(\Gamma )$ is a differentiable orbifold.

\subsection*{\normalsize{\bf 2.5. The space of 1-skeletons of Regge polytopes.}}
Let $Aut_{\partial }(P_{l})\subset Aut(P_{l})$, denote the subgroup of ribbon
graph automorphisms of the (trivalent) $1$-skeleton $\Gamma $ of $
|P_{T_{l}}|\rightarrow {M}$ that preserve the (labeling of the) boundary
components of $\Gamma $. Then, the space $K_{1}RP_{g,N_{0}}^{met}$ of $1$
-skeletons of conical Regge polytopes $|P_{T_{l}}|\rightarrow {M}$, with $
N_{0}(T)$ labelled boundary components, on a surface $M$ of genus $g$ can be
defined by \cite{mulase} 
\begin{equation}
K_{1}RP_{g,N_{0}}^{met}=\bigsqcup_{\Gamma \in RGB_{g,N_{0}}}\frac{\mathbb{R}
_{+}^{|e(\Gamma )|}}{Aut_{\partial }(P_{l})},  \label{DTorb}
\end{equation}
where the disjoint union is over the subset of all trivalent ribbon graphs
(with labelled boundaries) satisfying the topological stability condition $
2-2g-N_{0}(T)<0$, and which are dual to generalized triangulations. It
follows, (see \cite{mulase} theorems 3.3, 3.4, and 3.5), that the set $
K_{1}RP_{g,N_{0}}^{met}$ is locally modelled on a stratified space
constructed from the components (rational orbicells) $\mathbb{R}
_{+}^{|e(\Gamma )|}/Aut_{\partial }(P_{l})$ by means of a (Whitehead)
expansion and collapse procedure for ribbon graphs, which amounts to
collapsing edges and coalescing vertices, (the Whitehead move in $
|P_{T_{l}}|\rightarrow {M}$ is the dual of the familiar flip move for
triangulations). Explicitly, if $l(t)=tl$ is the length of an edge $\rho
^{1}(j)$ of a ribbon graph $\Gamma _{l(t)}\in $ $K_{1}RP_{g,N_{0}}^{met}$,
then, as $t\rightarrow 0$, we get the metric ribbon graph $\widehat{\Gamma }$
which is obtained from $\Gamma _{l(t)}$ by collapsing the edge $\rho ^{1}(j)$
. By exploiting such construction, we can extend the space $
K_{1}RP_{g,N_{0}}^{met}$ to a suitable closure $\overline{K_{1}RP}
_{g,N_{0}}^{met}$ \cite{loojenga}, (this natural topology on $K_{1}RP_{g,N_{0}}^{met}$
shows that, at least in two-dimensional quantum gravity, the set of Regge
triangulations with \emph{fixed connectivity} does not explore the full
configurational space of the theory). The open cells of 
$K_{1}RP_{g,N_{0}}^{met}$, being associated with trivalent graphs, have
dimension provided by the number $N_{1}(T)$ of edges of $|P_{T_{l}}|
\rightarrow {M}$. From the Euler relation $N_{0}(T)-N_{1}(T)+N_{2}(T)=2-2g$,
and the constraint $2N_{1}(T)=3N_{2}(T)$ associated with the trivalency, we
get 
\begin{equation}
\dim \left[ K_{1}RP_{g,N_{0}}^{met}\right] =N_{1}(T)=3N_{0}(T)+6g-6.
\end{equation}
There is a natural projection 
\begin{gather}
p:K_{1}RP_{g,N_{0}}^{met}\longrightarrow \mathbb{R}_{+}^{N_{0}(T)} \\
\Gamma \longmapsto p(\Gamma )=(l_{1},...,l_{N_{0}(T)}),  \notag
\end{gather}
where $(l_{1},...,l_{N_{0}(T)})$ denote the perimeters of the polygonal
2-cells $\{\rho ^{2}(j)\}$ of $|P_{T_{l}}|\rightarrow {M}$. With respect to
the topology on the space of metric ribbon graphs, the orbifold $
K_{1}RP_{g,N_{0}}^{met}$ endowed with such a projection acquires the
structure of a cellular bundle. For a given sequence $\{l(\partial (\rho
^{2}(k)))\}$, the fiber 
\begin{equation}
p^{-1}(\{l(\partial (\rho ^{2}(k)))\})=\left\{ |P_{T_{l}}|\rightarrow {M}\in
K_{1}RP_{g,N_{0}}^{met}:\{l_{k}\}=\{l(\partial (\rho ^{2}(k)))\}\right\}
\end{equation}
is the set of all generalized conical Regge polytopes with the given set of
perimeters. If we take into account the $N_{0}(T)$ constraints associated
with the perimeters assignments, it follows that the fibers $
p^{-1}(\{l(\partial (\rho ^{2}(k)))\})$ have dimension provided by 
\begin{equation}
\dim \left[ p^{-1}(\{l(\partial (\rho ^{2}(k)))\}\right] =2N_{0}(T)+6g-6,
\end{equation}
which exactly corresponds to the real dimension of the moduli space $\mathfrak{M}
_{g},_{N_{0}}$ of genus $g$ Riemann surfaces $((M;N_{0}),\mathcal{C})$ with 
$N_{0}$ punctures.

\subsection*{\normalsize{\bf 2.6. Orbifold labelling and dynamical triangulations.}}
Let us denote by 
\begin{equation}
\Omega _{T_{a}}\doteq \frac{\mathbb{R}_{+}^{|e(\Gamma )|}}{Aut_{\partial
}(P_{T_{a}})}  \label{omega}
\end{equation}
the rational cell associated with the 1-skeleton of the conical polytope $
|P_{T_{a}}|\rightarrow {M}$ dual to a dynamical triangulation $
|T_{l=a}|\rightarrow M$. The orbicell (\ref{omega}) contains the ribbon
graph associated with $|P_{T_{a}}|\rightarrow {M}$ and all (trivalent)
metric ribbon graphs $|P_{T_{L}}|\rightarrow {M}$ with the same
combinatorial structure of $|P_{T_{a}}|\rightarrow {M}$ but with all
possible length assignments $\{l(\rho ^{1}(h))\}_{1}^{N_{1}(T)}$ associated
with the corresponding set of edges $\{\rho ^{1}(h)\}_{1}^{N_{1}(T)}$. The
orbicell $\Omega _{T_{a}}$ is naturally identified with the convex
polytope (of dimension $(2N_{0}(T)+6g-6)$) in $\mathbb{R}_{+}^{N_{1}(T)}$
defined by 
\begin{equation}\label{strata}
\left\{ \{l(\rho ^{1}(j))\}\in \mathbb{R}_{+}^{N_{1}(T)}:
\sum_{j=1}^{q(k)}A_{(k)}^{j}(T_{a})L(\rho ^{1}(j))\,=\frac{\sqrt{3}}{3}
aq(k),\;k=1,...,N_{0}\;\right\} ,
\end{equation}
where $A_{(k)}^{j}(T_{a})$ is a $(0,1)$ indicator matrix, depending on the
given dynamical triangulation $|T_{l=a}|\rightarrow M$, with $
A_{(k)}^{j}(T_{a})=1$ if the edge $\rho ^{1}(j)$ belongs to $\partial (\rho
^{2}(k))$, and $0$ otherwise, and $\frac{\sqrt{3}}{3}aq(k)$ is the perimeter
length $l(\partial (\rho ^{2}(k)))$ in terms of the corresponding curvature
assignment $q(k)$. Note that $|P_{T_{a}}|\rightarrow {M}$ appears as the
barycenter of such a polytope.

Since the cell decomposition (\ref{DTorb}) of the space of trivalent metric
ribbon graphs $K_{1}RP_{g,N_{0}}^{met}$ depends only on the combinatorial
type of the ribbon graph, we can use the equilateral polytopes $
|P_{T_{a}}|\rightarrow {M}$, dual to dynamical triangulations, as the set
over which the disjoint union in (\ref{DTorb}) runs. Thus we can write 
\begin{equation}
K_{1}RP_{g,N_{0}}^{met}=\bigsqcup_{\mathcal{DT}(N_{0})}\Omega _{T_{a}},
\end{equation}
\ \ where 
\begin{equation}
\mathcal{DT}_{g}\left( N_{0}\right) \doteq \left\{ |T_{l=a}|\rightarrow
M\;:(\sigma ^{0}(k))\;k=1,...,N_{0}(T)\right\}
\end{equation}
denote the set of distinct generalized dynamically triangulated surfaces
of genus $g$, with a given set of $N_{0}(T)$ ordered labelled vertices.  

Note that \cite{mulase2}, even if the set $\mathcal{DT}_{g}\left( N_{0}\right) $ can be
considered (through barycentrical dualization) a well-defined subset of $
K_{1}RP_{g,N_{0}}^{met}$, it is not an orbifold over $\mathbb{N}$. For this
latter reason, the analysis of the metric stuctures over (generalized)
polytopes requires the use of the full orbicells $\Omega _{T_{a}}$ and we
cannot limit our discussion to equilateral polytopes.

\section{Some properties of the moduli space.}

At this point, it is worthwhile to discuss a few general aspects of moduli
space theory which come up in our geometrical analysis of triangulated
surfaces. We will do no more than summarize a few basic results and discuss
in some detail only the aspects of the theory which are of particular
relevance to us. We start by recalling that the moduli space $\mathfrak{M}
_{g},_{N_{0}}$ of genus $g$ Riemann surfaces $((M;N_{0}),\mathcal{C})$ with 
$N_{0}$ punctures is a dense open subset of a natural compactification
(Knudsen-Deligne-Mumford ) in a connected, compact orbifold of complex dimension $3g-3+N_0$ denoted
by $\overline{\mathfrak{M}}_{g},_{N_{0}}$. This latter is, by definition, the
moduli space of stable $N_{0}$-pointed curves of genus $g$, where a stable
curve is a compact Riemann surface with at most ordinary double points such
that its parts are hyperbolic. The closure $\partial \mathfrak{M}_{g},_{N_{0}}$
of $\mathfrak{M}_{g},_{N_{0}}$ in $\overline{\mathfrak{M}}_{g},_{N_{0}}$ consists of
stable curves with double points, and gives rise to a stratification
decomposing $\overline{\mathfrak{M}}_{g},_{N_{0}}$ into subvarieties. By
definition, a stratum of codimension $k$ is the component of $\overline{
\mathfrak{M}}_{g},_{N_{0}}$ parametrizing stable curves (of fixed topological
type) with $k$ double points.

\bigskip

\subsection*{\normalsize{\bf 3.1. Surface bundles and their canonical sections.}}
A basic observation in moduli space theory is the fact that any point $p$ on a
stable curve $((M;N_{0}),\mathcal{C})\in \overline{\mathfrak{M}}_{g},_{N_{0}}$
defines a natural mapping 
\begin{equation}
((M;N_{0}),\mathcal{C})\longrightarrow \overline{\mathfrak{M}}_{g},_{N_{0}+1}
\label{natmap}
\end{equation}
that determines a stable curve $((M;N_{0}+1),\mathcal{C}^{\prime })\in 
\overline{\mathfrak{M}}_{g},_{N_{0}+1}$. Explicitly, as long as the point $p$ is
disjoint from the puncture set $\{p_{k}\}_{k=1}^{N_{0}}$ one simply defines $
((M;N_{0}+1),\mathcal{C}^{\prime })$ to be $((M;N_{0},\{p\}),\mathcal{C})$.
If the point $p=p_{h}$ for some $p_{h}\in \{p_{k}\}_{k=1}^{N_{0}}$, then: 
\emph{\ i)} for any $1\leq i\leq N_{0}$, with $i\neq h$, identify $
p_{i}^{\prime }\in ((M;N_{0}+1),\mathcal{C}^{\prime })$ with the
corresponding $p_{i}$; \emph{ii)} take a three punctured sphere $\mathbb{CP}
_{(0,1,\infty )}^{1}$, label with a sub-index $h$ one of its punctures $
(0,1,\infty )$, say $\infty _{h}$, and attach it to the given $p_{h}\in 
\left[ ((M;N_{0}),\mathcal{C})\right] $; $\emph{iii)}$ relabel the remaining
two punctures $(0,1)\in \mathbb{CP}_{(0,1,\infty )}^{1}$ as $p_{h}^{\prime }$
and $p_{N_{0}+1}^{\prime }$. In this way, we get a genus $g$\ noded surface 
\begin{equation}
s_{h}\left[ ((M;N_{0}),\mathcal{C})\right] =\left[ ((M;N_{0}+1),\mathcal{C}
^{\prime })\right] \doteq ((M;N_{0}),\mathcal{C})_{_{h}}\cup \mathbb{CP}
_{(0,1,\infty _{h})}^{1}  \label{tail}
\end{equation}
with a rational tail and with a double point corresponding to the original
puncture $p_{h}$. Finally if $p$ happens to coincide with a node, then $
\left[ ((M;N_{0}+1),\mathcal{C}^{\prime })\right] $ results from setting $
p_{j}^{\prime }\doteq p_{j}$ for any $1\leq i\leq N_{0}$ and by: \emph{i)}
normalizing $\left[ ((M;N_{0}),\mathcal{C})\right] $ at the node (\emph{
i.e.}, by separating the branches of $\left[ ((M;N_{0}),\mathcal{C})\right] $
at $p$); \emph{ii)} inserting a copy of $\mathbb{CP}_{(0,1,\infty )}^{1}$
with $\{0,\infty \}$ identified with the preimage of $p$ and with $
p_{N_{0}+1}^{\prime }\doteq 1\in \mathbb{CP}_{(0,1,\infty )}^{1}$.
Conversely, let 
\begin{gather}
\pi :\overline{\mathfrak{M}}_{g},_{N_{0}+1}\longrightarrow \overline{\mathfrak{M}}
_{g},_{N_{0}} \\
\left[ ((M;N_{0}+1),\mathcal{C})\right] \vdash 
\begin{tabular}{c}
$forget$ \\ 
$\&$ \\ 
$collapse$
\end{tabular}
\longrightarrow \left[ ((M;N_{0}),\mathcal{C}^{\prime })\right]  \notag
\end{gather}
the projection which forgets the $(N_{0}+1)^{st}$ puncture and collapse to a
point any irreducible unstable component of the resulting curve. The fiber
of $\pi $ over $((M;N_{0}),\mathcal{C})$ is parametrized by the map (\ref
{natmap}), and if $((M;N_{0}),\mathcal{C})$ has a trivial automorphism group 
$\ Aut[((M;N_{0}),\mathcal{C})]\ $then $\pi ^{-1}((M;N_{0}),\mathcal{C})$ is
by definition the surface $((M;N_{0}),\mathcal{C})$, otherwise it is
identified with the quotient $((M;N_{0}),\mathcal{C})/Aut[((M;N_{0}),
\mathcal{C})]$. Thus, under the action of $\pi $, we can consider $
\overline{\mathfrak{M}}_{g},_{N_{0}+1}$ as a family (in the orbifold sense) of
Riemann surfaces over $\overline{\mathfrak{M}}_{g},_{N_{0}}$ and we can
identify $\overline{\mathfrak{M}}_{g},_{N_{0}+1}$ with the universal curve $
\overline{\mathcal{C}}_{g},_{N_{0}}$, 
\begin{equation}
\pi :\overline{\mathcal{C}}_{g},_{N_{0}}\longrightarrow \overline{\mathfrak{M}}
_{g},_{N_{0}}.
\end{equation}
Note that, by construction, $\overline{\mathcal{C}}_{g},_{N_{0}}$ (but for
our purposes is more profitable to think in terms of $\overline{\mathfrak{M}}
_{g},_{N_{0}+1}$) comes endowed with the $N_{0}$ natural sections $
s_{1},...,s_{N_{0}}$ 
\begin{gather}
s_{h}:\overline{\mathfrak{M}}_{g},_{N_{0}}\longrightarrow \overline{\mathcal{C}}
_{g},_{N_{0}} \\
\left[ ((M;N_{0}),\mathcal{C})\right] \longmapsto s_{h}\left[ ((M;N_{0}),
\mathcal{C})\right] \doteq ((M;N_{0}),\mathcal{C})_{_{h}}\cup \mathbb{CP}
_{(0,1,\infty _{h})}^{1},  \notag
\end{gather}
defined by (\ref{tail}).

\subsection*{\normalsize{\bf 3.2. The relative dualizing sheaf.}} 
The images of the sections $s_{i}$ characterize a divisor $\{D_{i}\}_{i=1}^{N_{0}}$ in $
\overline{\mathcal{C}}_{g},_{N_{0}}$ which has a great geometric interest
both in quantum gravity (where it is associated with the generation of 
MinBUs: Minimal Bottleneck Universes, the configurations characterizing the
susceptibility exponent of 2D gravity), and in discussing the topology of $\overline{\mathfrak{M}}_{g},_{N_{0}}$.  In both cases, such a study exploits
the properties of the tautological classes over $\overline{\mathcal{C}}
_{g},_{N_{0}}$ generated by the sections $\{s_{i}\}_{i=1}^{N_{0}}$ and by
the corresponding divisors $\{D_{i}\}_{i=1}^{N_{0}}$. To define such
classes, recall that the cotangent bundle (in the orbifold sense) to the
fibers of of the universal curve $\pi :\overline{\mathcal{C}}
_{g},_{N_{0}}\longrightarrow \overline{\mathfrak{M}}_{g},_{N_{0}}$ gives rise to
a holomorphic line bundle $\omega _{g,N_{0}}\doteq \omega _{\overline{
\mathcal{C}}_{g},_{N_{0}}/\overline{\mathfrak{M}}_{g},_{N_{0}}}$ over $
\overline{\mathcal{C}}_{g},_{N_{0}}$ (the relative dualizing sheaf of $\pi :
\overline{\mathcal{C}}_{g},_{N_{0}}\longrightarrow \overline{\mathfrak{M}}
_{g},_{N_{0}}$), this is essentially the sheaf of 1-forms with a
natural polar behavior along the possible nodes of the Riemann surface
describing the fiber of $\pi $. A more explicit characterization, however,
will be needed later on, so we briefly pause to describe it here. In
particular, we will be interested on the behavior of the relative dualizing
sheaf $\omega _{g,N_{0}}$\ restricted to the generic divisor $D_{h}$
generated by the section $s_{h}$. To this end, let $z_{1}(h)$ and $
z_{2}(\infty _{h})$ denote local coordinates defined in the disks $\Delta
_{p_{h}}\doteq \{|z_{1}(h)|<1\}$ and $\Delta _{\infty _{h}}\doteq
\{|z_{2}(\infty _{h})|<1\}$ respectively centered around the punctures $
p_{h}\in ((M;N_{0}),\mathcal{C})$, and $\infty \in \mathbb{CP}_{(0,1,\infty
)}^{1}$. \ Let $\Delta _{t_{h}}=\{t_{h}\in \mathbb{C}:|t_{h}|<1\}$. Consider
the analytic family $s_{h}(t_{h})$ of surfaces of genus $g$ defined over $
\Delta _{t_{h}}$ and obtained by removing the disks $|z_{1}(h)|<|t_{h}|$ and 
$|z_{2}(\infty _{h})|<|t_{h}|$ from $((M;N_{0}),\mathcal{C})$ and $\mathbb{CP
}_{(0,1,\infty )}^{1}$ and gluing the resulting surfaces through the annulus 
$\{(z_{1}(h),z_{2}(\infty _{h}))|\;z_{1}(h)z_{2}(\infty
_{h})=t_{h},\,t_{h}\in \Delta _{t_{h}}\}$\ by identifying the points of
coordinate $z_{1}(h)$ with the points of coordinates $z_{2}(\infty
_{h})=t_{h}/z_{1}(h)$. The family $s_{h}(t_{h})\rightarrow \Delta _{t_{h}}$
opens the node $z_{1}(h)z_{2}(\infty _{h})=0$ of the section $
s_{h}|_{((M;N_{0}),\mathcal{C})}$. Note that in such a way we can
independently and holomorphically open the distinct nodes of the various
sections $\{s_{k}\}_{k=1}^{N_0}$. More generally, while opening the node we
can also vary the complex structure of $((M;N_{0}),\mathcal{C})$ by
introducing local complex coordinates $(\tau _{\alpha })_{\alpha
=1}^{3g-3+N_{0}}$ for $\mathfrak{M}_{g},_{N_{0}}$ around $((M;N_{0}),\mathcal{C}
) $. If 
\begin{equation}
s_{h}(\tau _{\alpha },t_{h})\rightarrow \mathfrak{M}_{g},_{N_{0}}\times \Delta
_{t_{h}}  \label{covering}
\end{equation}
denotes the family of surfaces opening of the node , then in the
corresponding coordinates $(\tau _{\alpha },t_{h})$ the divisor $D_{h}$,
image of the section $s_{h}$, is locally defined by the equation $t_{h}=0$.
Similarly, the divisor $D\doteq \sum_{h=1}^{N_{0}}D_{h}$ is characterized by
the locus of equation $\prod_{h=1}^{N_{0}}t_{h}=0$.

\bigskip

The elements of the dualizing sheaf $\omega
_{g,N_{0}}|_{s_{h}(t_{h})}\doteq \omega _{g,N_{0}}(D_{h})$ are differential
forms $u(h)=u_{1}dz_{1}(h)+u_{2}dz_{2}(\infty _{h})$ such that $u(h)\wedge
dt_{h}=fdz_{1}(h)\wedge dz_{2}(\infty _{h})$, where $f$ is a holomorphic
function of $z_{1}(h)$ and $z_{2}(\infty _{h})$. By differentiating $
z_{1}(h)z_{2}(\infty _{h})=t_{h}$, one gets $f=u_{1}z_{1}(h)-u_{2}z_{2}(
\infty _{h})$ which is the defining relation for the forms in $\omega
_{g,N_{0}}(D_{h})$. In particular, by choosing $u_{1}=f/2z_{1}(h)$, and $
u_{2}=f/2z_{2}(\infty _{h})$ we get the local isomorphism between the sheaf
of holomorphic functions $\mathcal{O}_{s_{h}(t_{h})}$ over $s_{h}(t_{h})\ $
and $\omega _{g,N_{0}}(D_{h})$
\begin{equation}
f\longmapsto u(h)=f\left( \frac{1}{2}\frac{dz_{1}(h)}{z_{1}(h)}-\frac{1}{2}
\frac{dz_{2}(\infty _{h})}{z_{2}(\infty _{h})}\right) .
\end{equation}
If we set $f=f_{0}+f_{1}(z_{1}(h))+f_{2}(z_{2}(\infty _{h}))$, where $f_{0}$
is a constant and $f_{1}(0)=0=f_{2}(0)$, then on the noded surface $s_{h}$ ,
($t_{h}=0$), we get from the relation $z_{2}(\infty
_{h})dz_{1}(h)+z_{1}(h)dz_{2}(\infty _{h})=0$, 
\begin{eqnarray}
u_{h}|_{z_{2}(\infty _{h})=0} &=&\frac{f_{0}+f_{1}(z_{1}(h))}{z_{1}(h)}
dz_{1}(h), \\
u_{h}|_{z_{1}(h)=0} &=&-\frac{f_{0}+f_{2}(z_{2}(\infty _{h}))}{z_{2}(\infty
_{h})}dz_{2}(\infty _{h}),  \notag
\end{eqnarray}
on the two branches $\Delta _{p_{h}}\cap ((M;N_{0}),\mathcal{C})$ and $
\Delta _{\infty _{h}}\cap \mathbb{CP}_{(0,1,\infty )}^{1}$ of the node where 
$z_{1}(h)$ and $z_{2}(\infty _{h})$ are a local coordinate (\emph{i.e.} $
z_{2}(\infty _{h})=0$ and $z_{1}(h)=0$, respectively). Thus, near the node
of $s_{h}$, $\omega _{g,N_{0}}(D_{h})\ $is generated by $\frac{dz_{1}(h)}{
z_{1}(h)}$ and $\frac{dz_{2}(\infty _{h})}{z_{2}(\infty _{h})}$ subjected to
the relation $\frac{dz_{1}(h)}{z_{1}(h)}+\frac{dz_{2}(\infty _{h})}{
z_{2}(\infty _{h})}=0$. Stated differently, a section of the sheaf $\omega
_{g,N_{0}}(D_{h})\ $pulled back to the smooth normalization $((M;N_{0}),
\mathcal{C})_{p_{h}}\bigsqcup \mathbb{CP}_{(0,1,\infty _{h})}^{1}$ of $s_{h}$
can be identified with a meromorphic $1$-form with at most simple poles at
the punctures $p_{h}$ and $\infty _{h}$ which are identified under the
normalization map, and with opposite residues at such punctures. By
extending such a construction to all $N_{0}$ sections $\{s_{h}
\}_{h=1}^{N_{0}}$, we can define the line bundle 
\begin{equation}
\omega _{g,N_{0}}(D)\doteq \omega _{g,N_{0}}\left(
\sum_{i=1}^{N_{0}}D_{i}\right) \longrightarrow \overline{\mathcal{C}}
_{g},_{N_{0}}
\end{equation}
as $\omega _{g,N_{0}}$ twisted by the divisor $D\doteq
\sum_{h=1}^{N_{0}}D_{h}$, \emph{viz}. the line bundle locally generated by
the differentials $\frac{dz_{1}(h)}{z_{1}(h)}$ for $z_{1}(h)\neq 0$ and $-$ $
\frac{dz_{2}(\infty _{h})}{z_{2}(\infty _{h})}$ for $z_{2}(\infty _{h})\neq
0 $, with $z_{1}(h)z_{2}(\infty _{h})=0$, and $h=1,...,N_{0}$. As above, 
$\{z_{1}(h)\}_{h=1}^{N_{0}}$ are local variables at the marked points $
\{p_{h}\}_{i=1}^{N_{0}}\in ((M;N_{0}),\mathcal{C})$, whereas $z_{2}(\infty
_{h})$ is the corresponding variable in the three punctured sphere $\mathbb{
CP}_{(0,1,\infty )}^{1}$.

\bigskip

\subsection*{\normalsize{\bf 3.3. Meromorphic quadratic differentials and their
Weil-Petersson norm.}} 
We can associate with the family of surfaces $s_{h}(\tau _{\alpha },t_{h})$, opening the node $t_{h}=0$ of the section 
$s_{h}(\tau _{\alpha })$, a meromorphic quadratic differential which, in
terms of the local covering for $\mathfrak{M}_{g},_{N_{0}}$ defined by (\ref
{covering}), is given by \cite{wolpert}
\begin{equation}
\Phi (h;\tau _{\alpha },t_{h})\doteq \left\{ 
\begin{tabular}{ccc}
$-\frac{t_{h}}{\pi }\frac{dz_{1}(h)\otimes dz_{1}(h)}{z_{1}(h)^{2}},$ & in & 
$\Delta _{p_{h}}\cap ((M;N_{0}),\mathcal{C})$ \\ 
$-\frac{t_{h}}{\pi }\frac{dz_{2}(\infty _{h})\otimes dz_{2}(\infty _{h})}{
z_{2}(\infty _{h})^{2}},$ & in & $\Delta _{\infty _{h}}\cap \mathbb{CP}
_{(0,1,\infty )}^{1}.$
\end{tabular}
\right.  \label{modquad}
\end{equation}
The quadratic differential $\Phi (h;\tau _{\alpha },t_{h})$ is an element of
the (relative) $2$-canonical bundle $\omega _{g,N_{0}}(D_{h})^{\otimes 2}$ 
originating from the basis $\frac{dz_{1}(h)}{z_{1}(h)}$ of $\omega
_{g,N_{0}}(D_{h})$, so, it can be thought of as a cotangent vector to $\mathfrak{
M}_{g},_{N_{0}}$, dual to the holomorphic vector field $\frac{\partial }{
\partial t_{h}}$. In any annulus $\{|t_{h}|<|z_{1}(h)|<1\}$ parametrizing
the opening of the node family $s_{h}(\tau _{\alpha },t_{h})$, we can
evaluate the Weil-Petersson norm \cite{wolpert} of the quadratic
differential $\Phi (h;\tau _{\alpha },t_{h})$ according to 
\begin{equation}
\left\| \Phi (h;\tau _{\alpha },t_{h})\right\|
_{W-P}=\int_{\{|t_{h}|<|z_{1}(h)|<1\}}\frac{|\Phi (h;\tau _{\alpha
},t_{h})|^{2}}{g_{hyp}},
\end{equation}
where $g_{hyp}$ denotes the hyperbolic metric 
\begin{equation}
g_{hyp}\doteq \left( \frac{\pi ^{2}}{\ln ^{2}|t_{h}|}\right) \frac{dz_{1}(h)d
\overline{z_{1}(h)}}{|z_{1}(h)|^{2}\sin ^{2}\left( \pi \frac{\ln |z_{1}(h)|}{
\ln |t_{h}|}\right) }
\end{equation}
on $\{|t_{h}|<|z_{1}(h)|<1\}$. A direct computation provides \cite{wolpert} 
\begin{equation}
\left\| \Phi (h;\tau _{\alpha },t_{h})\right\| _{W-P}=\left| t_{h}\right|
^{2}\left( \frac{1}{\pi }\ln \frac{1}{|t_{h}|}\right) ^{3}.
\end{equation}
The corresponding Weil-Petersson metric $ds_{W-P}^{2}(h)$ and the associated
K\"{a}hler form $\omega _{W-P}(h)$ are respectively provided by 
\begin{equation}
ds_{W-P}^{2}(h)=\frac{2\pi ^{3}\left| dt_{h}\right| ^{2}}{\left|
t_{h}\right| ^{2}\left( \ln \frac{1}{|t_{h}|}\right) ^{3}},
\end{equation}
and 
\begin{equation}
\omega _{W-P}(h)=\sqrt{-1}\frac{2\pi ^{3}}{\left| t_{h}\right| ^{2}\left(
\ln \frac{1}{|t_{h}|}\right) ^{3}}dt_{h}\wedge d\overline{t_{h}}.
\end{equation}
Observe here a similarity between the above definitions and properties of
the sections $\{s_{h}\}$ and the construction of minimal bottleneck
universes in simplicial quantum gravity. This is not a mere coincidence, and
it will turn out that such a similarity has a far reaching role in
discussing the complex analytic geometry of Regge polytopes and its
connection with 2D simplicial quantum gravity.

\bigskip

\subsection*{\normalsize{\bf 3.4. The ribbon graph parametrization of the moduli space.}} 
The complex analytic geometry of the space of conical Regge
polytopes which we will discuss in the next section generalizes the
well-known bijection (a homeomorphism of orbifolds) between the space of
metric ribbon graphs $K_{1}RP_{g,N_{0}}^{met}$ (which forgets the conical
geometry) and the moduli space $\mathfrak{M}_{g},_{N_{0}}$ of genus $g$ Riemann
surfaces $((M;N_{0}),\mathcal{C})$ with $N_{0}(T)$\ punctures \cite{mulase}, \cite{loojenga}. 
This bijection results in a local parametrization of $\mathfrak{M}_{g},_{N_{0}}$
defined by 
\begin{gather}
h:K_{1}RP_{g,N_{0}}^{met}\rightarrow \mathfrak{M}_{g},_{N_{0}}\times {R}_{+}^{N}
\label{bijec} \\
\Gamma \longmapsto \lbrack ((M;N_{0}),\mathcal{C}),l_{i}]  \notag
\end{gather}
where $(l_{1},...,l_{N_{0}})$ is an ordered n-tuple of positive real numbers
and $\Gamma $ is a metric ribbon graph with $N_{0}(T)$ labelled boundary
lengths $\{l_{i}\}$ . 
If $\overline{K_{1}RP}_{g,N_{0}}^{met}$ is the closure
of $K_{1}RP_{g,N_{0}}^{met}$, then the bijection $h$ extends to $\overline{
K_{1}RP}_{g,N_{0}}^{met}\rightarrow \overline{\mathfrak{M}}_{g},_{N_{0}}\times {R
}_{+}^{N_{0}}$ in such a way that a ribbon graph $\Gamma \in \overline{RGP}
_{g,N_{0}}^{met}$ is mapped in two (stable) surfaces $M_{1}$ and $M_{2}$
with $N_{0}(T)$ punctures if and only if there exists an homeomorphism
between $M_{1}$ and $M_{2}$ preserving the (labelling of the) punctures, and
is holomorphic on each irreducible component containing one of the punctures.

\begin{figure}[ht]
\begin{center}
\includegraphics[scale=.4]{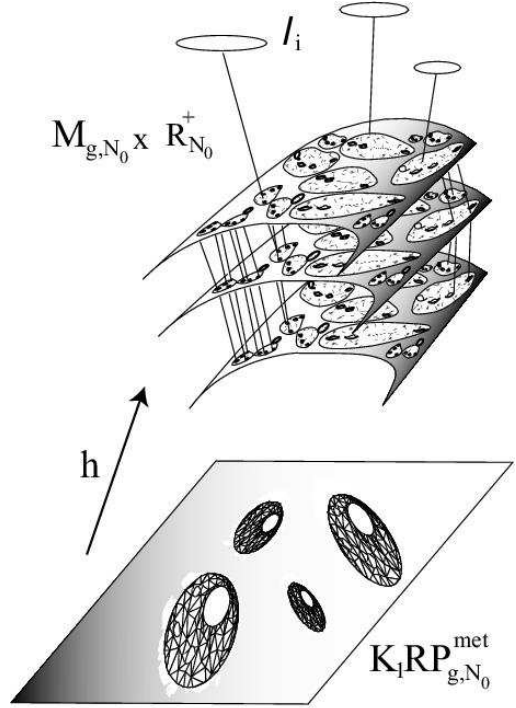}
\caption{The map $h$ associates to each ribbon graph an element of the standard moduli space $\mathfrak{M}_{g,N_{0}}\times\Re^{+}_{N_{0}}$.}
\end{center}
\end{figure}

According to Kontsevich \cite{kontsevich}, corresponding to each marked polygonal 2-cells $
\{\rho ^{2}(k)\}$ of $|P_{T_{l}}|\rightarrow {M}$ there is a further
(combinatorial) bundle map 
\begin{equation}
\mathcal{CL}_{k}\rightarrow K_{1}RP_{g,N_{0}}^{met}  \label{combundle}
\end{equation}
whose fiber over $(\Gamma ,\rho ^{2}(1),...,\rho ^{2}(N_{0}))$ is provided
by the boundary cycle $\partial \rho ^{2}(k)$, (recall that each boundary $
\partial \rho ^{2}(k)$ comes with a positive orientation).\\ 
To any such cycle one associates \cite{brezin}, \cite{loojenga} the corresponding perimeter map $l(\partial (\rho
^{2}(k)))=\sum $ $l(\rho ^{1}(h_{\alpha }))$ which then appears as defining
a natural connection on $\mathcal{CL}_{k}$. The piecewise smooth 2-form
defining the curvature of such a connection, 
\begin{equation}
\omega _{k}(\Gamma )=\sum_{1\leq h_{\alpha }<h_{\beta }\leq q(k)-1}d\left( 
\frac{l(\rho ^{1}(h_{\alpha }))}{l(\partial \rho ^{2}(k))}\right) \wedge
d\left( \frac{l(\rho ^{1}(h_{\beta }))}{l(\partial \rho ^{2}(k))}\right) ,
\label{chern}
\end{equation}
is invariant under rescaling and cyclic permutations of the $l(\rho
^{1}(h_{\mu }))$, and is a combinatorial representative of the Witten class
of the line bundle $\mathcal{L}_{k}$.

\bigskip

It is important to stress that even if ribbon graphs can be thought of as
arising from Regge polytopes (with variable connectivity), (\ref{bijec})
only involves the ribbon graph structure and the theory can be (and actually
is) developed with no reference at all to a particular underlying
triangulation. In such a connection, the role of dynamical triangulations
has been slightly overemphasized, they simply provide a convenient way of
labelling the different combinatorial strata of the mapping (\ref{bijec}),
but, by themselves they do not define a combinatorial parametrization of $
\overline{\mathfrak{M}}_{g},_{N_{0}}$ for any finite $N_{0}$. However, it is
very useful, at least for the purposes of quantum gravity, to remember the
possible genesis of a ribbon graph from an underlying triangulation and be
able to exploit the further information coming from the associated conical
geometry. Such an information cannot be recovered from the ribbon graph
itself (with the notable exception of equilateral ribbon graphs, which can
be associated with dynamical triangulations), and must be suitably codified
by adding to the boundary lengths $\{l_{i}\}$ of the graph a further
decoration. This can be easily done by explicitly connecting Regge polytopes
to punctures Riemann surfaces.

\noindent

\section{Regge polytopes and punctured Riemann surfaces.}

As suggested by (\ref{cmetr}), the polyhedral metric associated with the
vertices $\{\sigma ^{0}(i)\}$ of a (generalized) Regge triangulation $
|T_{l}|\rightarrow M$, can be conveniently described in terms of complex
function theory. We can associate with $|P_{T_{l}}|\rightarrow {M}$ a
complex structure $((M;N_{0}),\mathcal{C})$ (a punctured Riemann surface)
which is, in a well-defined sense, dual to the structure (\ref{cmetr})
generated by $|T_{l}|\rightarrow M$. Let $\rho ^{2}(k)$ be the generic
two-cell $\in |P_{T_{l}}|\rightarrow {M}$ barycentrically dual to the vertex 
$\sigma ^{0}(k)\in |T_{l}|\rightarrow M$ . To the generic edge $\rho
^{1}(h) $ of $\rho ^{2}(k)$ we associate a complex uniformizing
coordinate $z(h)$ defined in the strip 
\begin{equation}
U_{\rho ^{1}(h)}\doteq \{z(h)\in \mathbb{C}|0<\func{Re}z(h)<l(\rho
^{1}(h))\},
\end{equation}
$l(\rho ^{1}(h))$ being the length of the edge considered. The uniformizing
coordinate $w(j)$, corresponding to the generic $3$-valent vertex $\rho
^{0}(j)\in \rho ^{2}(k)$, is defined in the open set 
\begin{equation}
U_{\rho ^{0}(j)}\doteq \{w(j)\in \mathbb{C}|\;|w(j)|<\delta ,\;w(j)[\rho
^{0}(j)]=0\},
\end{equation}
where $\delta >0$ is a suitably small constant. Finally, the two-cell $\rho
^{2}(k)$ is uniformized in the unit disk 
\begin{equation}
U_{\rho ^{2}(k)}\doteq \{\zeta (k)\in \mathbb{C}|\;|\zeta (k)|<1,\;\zeta
(k)[\sigma ^{0}(k)]=0\},
\end{equation}
where $\sigma ^{0}(k)$ is the vertex $\in |T_{l}|\rightarrow M$ 
corresponding to the given two-cell.

The various uniformizations $\{w(j),U_{\rho ^{0}(j)}\}_{j=1}^{N_{2}(T)}$, 
$\{z(h),U_{\rho ^{1}(h)}\}_{h=1}^{N_{1}(T)}$, and $\{\zeta (k),U_{\rho
^{2}(k)}\}_{k=1}^{N_{0}(T)}$ can be coherently glued together by noting that
to each edge $\rho ^{1}(h)\in $ $\rho ^{2}(k)$\ we can associate the
standard quadratic differential on $U_{\rho ^{1}(h)}$ given by 
\begin{equation}
\phi (h)|_{\rho ^{1}(h)}=dz(h)\otimes dz(h).  \label{foliat}
\end{equation}
Such $\phi (h)|_{\rho ^{1}(h)}$ can be extended to the remaining local
uniformizations $U_{\rho ^{0}(j)}$, and $U_{\rho ^{2}(k)}$, by exploiting
a classic result in Riemann surface theory according to which a quadratic
differential $\phi $ has a finite number of zeros $n_{zeros}(\phi )$ with
orders $k_{i}$ and a finite number of poles $n_{poles}(\phi )$ of order $
s_{i}$ such that 
\begin{equation}
\sum_{i=1}^{n_{zero}(\phi )}k_{i}-\sum_{i=1}^{n_{pole}(\phi )}s_{i}=4g-4.
\label{quadrel}
\end{equation}
In our case we must have $n_{zeros}(\phi )=N_{2}(T)$ with $k_{i}=1$,
(corresponding to the fact that the $1$-skeleton of $|P_{l}|\rightarrow M$
is a trivalent graph), and $n_{poles}(\phi )=$ $N_{0}(T)$ with $
s_{i}=s\;\forall i$, for a suitable positive integer $s$. According to such
remarks (\ref{quadrel}) reduces to 
\begin{equation}
N_{2}(T)-sN_{0}(T)=4g-4.  \label{poles}
\end{equation}
From the Euler relation $N_{0}(T)-N_{1}(T)+N_{2}(T)=2-2g$, and $
2N_{1}(T)=3N_{2}(T)$ we get $N_{2}(T)-2N_{0}(T)=4g-4$. This is consistent
with (\ref{poles}) if and only if $s=2$. Thus the extension $\phi $ of $
\phi (h)|_{\rho ^{1}(h)}$ along the $1$-skeleton of $|P_{l}|\rightarrow M$
must have $N_{2}(T)$ zeros of order $1$ corresponding to the trivalent
vertices $\{\rho ^{0}(j)\}$ of $|P_{l}|\rightarrow M$ and $N_{0}(T)$
quadratic poles corresponding to the polygonal cells $\{\rho ^{2}(k)\}$ of
perimeter lengths $\{l(\partial (\rho ^{2}(k)))\}$.

\begin{figure}[ht]
\begin{center}
\includegraphics[scale=.4]{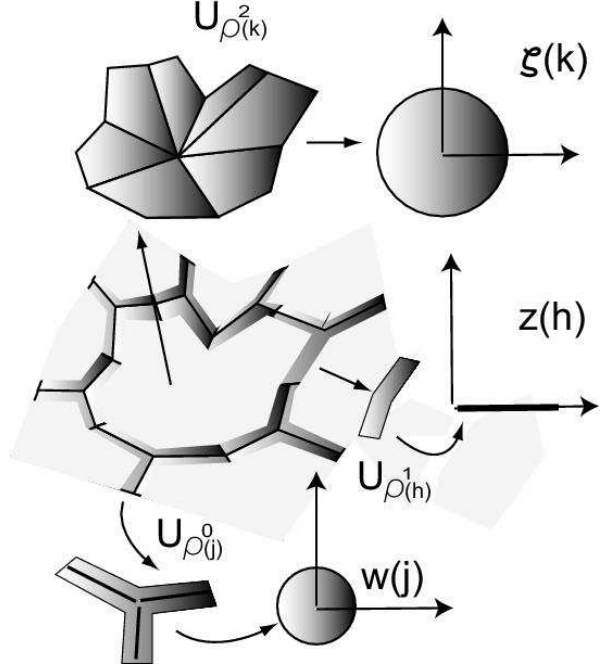}
\caption{The local presentation of uniformizing coordinates.}
\end{center}
\end{figure}

\noindent Around a zero of order one and a pole of order two, every (Jenkins-Strebel) quadratic differential 
$\phi$ has a canonical local structure which (along with (\ref{foliat})) is
given by \cite{mulase}, \cite{strebel} ,
\begin{equation}
(|P_{T_{l}}|\rightarrow {M)\rightarrow }\phi \doteq \left\{ 
\begin{tabular}{l}
$\phi (h)|_{\rho ^{1}(h)}=dz(h)\otimes dz(h),$ \\ 
$\phi (j)|_{\rho ^{0}(j)}=\frac{9}{4}w(j)dw(j)\otimes dw(j),$ \\ 
$\phi (k)|_{\rho ^{2}(k)}=-\frac{\left[ l(\partial (\rho ^{2}(k)))\right]
^{2}}{4\pi ^{2}\zeta ^{2}(k)}d\zeta (k)\otimes d\zeta (k),$
\end{tabular}
\right.  \label{differ}
\end{equation}
where $\{\rho ^{0}(j),\rho ^{1}(h),\rho ^{2}(k)\}$ runs over the set of
vertices, edges, and $2$-cells of $|P_{l}|\rightarrow M$. Since $\phi
(h)|_{\rho ^{1}(h)}$, $\phi (j)|_{\rho ^{0}(j)}$, and $\phi (k)|_{\rho
^{2}(k)}$ must be identified on the non-empty pairwise intersections $
U_{\rho ^{0}(j)}\cap U_{\rho ^{1}(h)}$, $U_{\rho ^{1}(h)}\cap U_{\rho
^{2}(k)}$ we can associate to the polytope $|P_{T_{l}}|\rightarrow {M}$ a
complex structure $((M;N_{0}),\mathcal{C})$ by coherently gluing, along the
pattern associated with the ribbon graph $\Gamma $, the local
uniformizations $\{U_{\rho ^{0}(j)}\}_{j=1}^{N_{2}(T)}$, $\{U_{\rho
^{1}(h)}\}_{h=1}^{N_{1}(T)}$, and $\{U_{\rho ^{2}(k)}\}_{k=1}^{N_{0}(T)}$.
Explicitly, let $\{U_{\rho ^{1}(j_{\alpha })}\}$, $\alpha =1,2,3$ be the
three generic open strips associated with the three cyclically oriented
edges $\{\rho ^{1}(j_{\alpha })\}$ incident on the generic vertex $\rho
^{0}(j)$. Then the uniformizing coordinates $\{z(j_{\alpha })\}$ are related
to $w(j)$ by the transition functions 
\begin{equation}
w(j)=e^{2\pi i\frac{\alpha -1}{3}}z(j_{\alpha })^{\frac{2}{3}},\hspace{0.2in}
\hspace{0.1in}\alpha =1,2,3.  \label{glue1}
\end{equation}
Note that in such uniformization the vertices $\{$ $\rho ^{0}(j)\}$ do not
support conical singularities since each strip $U_{\rho ^{1}(j_{\alpha })}$
is mapped by (\ref{glue1}) into a wedge of angular opening $\frac{2\pi }{3}$. 
This is consistent with the definition of $|P_{T_{l}}|\rightarrow {M}$
according to which the vertices $\{\rho ^{0}(j)\}\in |P_{T_{l}}|\rightarrow {
M}$ are the barycenters of $\ $the flat $\{\sigma ^{2}(j)\}\in
|T_{l}|\rightarrow M$. Similarly, if $\{U_{\rho ^{1}(k_{\beta })}\}$, $
\beta =1,2,...,q(k)$ are the open strips associated with the $q(k)$
(oriented) edges $\{\rho ^{1}(k_{\beta })\}$ boundary of the generic
polygonal cell $\rho ^{2}(k)$, then the transition functions between the
corresponding uniformizing coordinate $\zeta (k)$ and the $\{z(k_{\beta })\}$
are given by \cite{mulase} 
\begin{equation}
\zeta (k)=\exp \left( \frac{2\pi i}{l(\partial (\rho ^{2}(k)))}\left(
\sum_{\beta =1}^{\nu -1}l(\rho ^{1}(k_{\beta }))+z(k_{\nu })\right) \right) ,
\hspace{0.2in}\nu =1,...,q(k),  \label{glue2}
\end{equation}
with $\sum_{\beta =1}^{\nu -1}\cdot \doteq 0$, for $\nu =1$.

\bigskip

\subsection*{\normalsize{\bf 4.1. A Parametrization of the conical geometry.}} 
As for the metrical properties of the complex structure $((M;N_{0}),\mathcal{C})$
associated with $|P_{T_{l}}|\rightarrow {M}$ note that for any closed
curve $c:\mathbb{S}^{1}\rightarrow U_{\rho ^{2}(k)}$, homotopic to the
boundary of $\overline{U}_{\rho ^{2}(k)}$, we get 
\begin{equation}
\oint_{c}\sqrt{\phi (k)_{\rho ^{2}(k)}}=l(\partial (\rho ^{2}(k))).
\label{length}
\end{equation}
In general, let   
\begin{equation}
|\phi (k)_{\rho ^{2}(k)}|=\frac{\left[ l(\partial (\rho ^{2}(k)))\right] ^{2}
}{4\pi ^{2}|\zeta (k)|^{2}}|d\zeta (k)|^{2},  \label{flmetr}
\end{equation}
denote the usual cylindrical metric canonically associated with a quadratic
differential with a second order pole. If we define $\Delta _{k}^{\varrho
}\doteq \left\{ \zeta (k)\in \mathbb{C}|\;\varrho \leq |\zeta (k)|\leq
1\right\} $, then in terms of the area element associated with the flat
metric $|\phi (k)_{\rho ^{2}(k)}|$,(\emph{i.e.}, the absolute value of $
\phi (k)$), we get 
\begin{equation}
\int_{\Delta _{k}^{\varrho }}\frac{i}{2}\frac{\left[ l(\partial (\rho
^{2}(k)))\right] ^{2}}{4\pi ^{2}|\zeta (k)|^{2}}d\zeta (k)\wedge \ d
\overline{\zeta }(k)=\frac{\left[ l(\partial (\rho ^{2}(k)))\right] ^{2}}{
2\pi }\ln \left( \frac{1}{\varrho }\right) ,  \label{area}
\end{equation}
which, as $\varrho \rightarrow 0^{+}$, diverges logarithmically. Thus, as
already recalled (section 2.3) the punctured disk $\Delta _{k}^{\ast }$,
endowed with the flat metric $|\phi (k)_{\rho ^{2}(k)}|$, is
isometric to a flat semi-infinite cylinder. It is perhaps worthwhile
stressing that even if this latter geometry is perfectly consistent with the
metric ribbon graph structure, it is not really the natural metric to use if
we wish to explicitly keep track of the conical Regge polytope (and the
associated triangulation) which generate the given ribbon graph. For
instance, in the Kontsevich-Witten model, the ribbon graph structure and the
associated geometry $(\Delta _{k}^{\ast },|\phi (k)_{\rho ^{2}(k)}|)$ of 
the cells $U_{\rho ^{2}(k)}$ is sufficient to combinatorially describe
intersection theory on moduli space. However, a full-fledged study of 2D
simplicial quantum gravity (\emph{e.g.}, of Liouville theory) requires a
study of the full Regge geometry $(\Delta _{k}^{\ast },ds_{(k)}^{2})$. 

\begin{figure}[ht]
\begin{center}
\includegraphics[scale=.4]{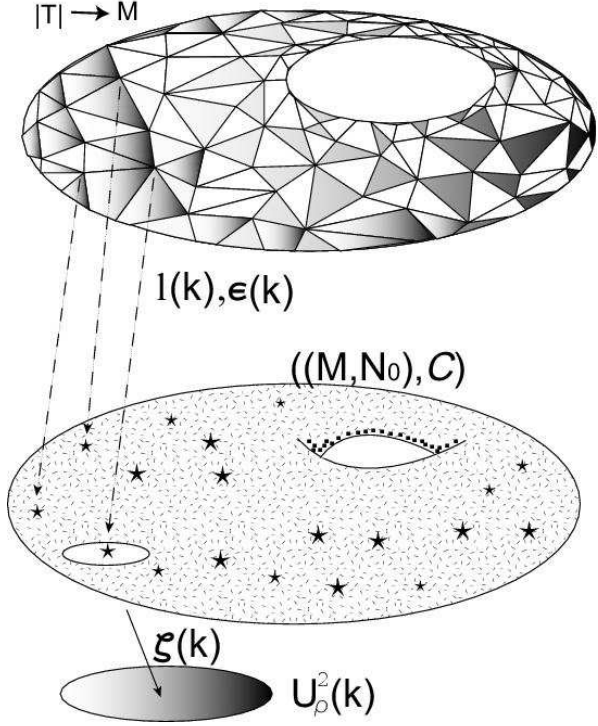}
\caption{The decorated Riemann surface associated with a Regge triangulation.}
\end{center}
\end{figure}

\noindent We can keep track of the conical geometry of the polygonal cell $\rho
^{2}(k)\in $ $|P_{T_{l}}|\rightarrow {M}$ by noticing that given a
normalized deficit angle $\frac{\varepsilon (k)}{2\pi }\doteq 1-\frac{\theta
(k)}{2\pi }$, the conical geometry (\ref{cmetr}) and the cylindrical
geometry (\ref{flmetr}) on the punctured disk $\Delta _{k}^{\ast }\subset
U_{\rho ^{2}(k)}$ 
\begin{equation}
\Delta _{k}^{\ast }\doteq \{\zeta (k)\in \mathbb{C}|\;0<|\zeta (k)|<1\}
\label{puncdisk}
\end{equation}
can be conformally related by choosing the conformal factor $e^{2u}$ in (\ref
{cmetr}) according to 
\begin{equation}
e^{2u}=\frac{\left[ l(\partial (\rho ^{2}(k)))\right] ^{2}}{4\pi ^{2}}\left|
\zeta (k)\right| ^{-2\left( \frac{\theta (k)}{2\pi }\right) }.
\end{equation}
Thus, the conical metric 
\begin{eqnarray}
ds_{(k)}^{2} &\doteq &\frac{\left[ l(\partial (\rho ^{2}(k)))\right] ^{2}}{
4\pi ^{2}}\left| \zeta (k)\right| ^{-2\left( \frac{\varepsilon (k)}{2\pi }
\right) }\left| d\zeta (k)\right| ^{2}=  \label{metrica} \\
&=&\frac{\left[ l(\partial (\rho ^{2}(k)))\right] ^{2}}{4\pi ^{2}}\left|
\zeta (k)\right| ^{2\left( \frac{\theta (k)}{2\pi }\right) }\frac{\left|
d\zeta (k)\right| ^{2}}{\left| \zeta (k)\right| ^{2}},  \notag
\end{eqnarray}
can be used to describe, as $\frac{\theta (k)}{2\pi }$ varies in $\mathbb{R}
^{+}$, all possible conical geometries on $\Delta _{k}^{\ast }\subset
U_{\rho ^{2}(k)}$ including the cylindrical metric $|\phi (k)_{\rho
^{2}(k)}| $ as a particular case corresponding to $\frac{\theta (k)}{2\pi }
=0 $.

\bigskip

Since $|\phi (k)_{\rho ^{2}(k)}|$ and $ds_{(k)}^{2}$ correspond to the
same moduli point $((M;N_{0}),\mathcal{C})$ $\in \mathfrak{M}_{g},_{N_{0}}$ we
can naturally extend the explicit construction \cite{mulase} of the mapping (\ref
{bijec}) for defining the decorated Riemann surface corresponding to a
conical Regge polytope:

\begin{proposition}\label{gluing}
Let $\{p_{k}\}_{k=1}^{N_{0}}\in M$ denote the set of punctures corresponding
to the decorated vertices $\{\sigma ^{0}(k),\frac{\varepsilon (k)}{2\pi }
\}_{k=1}^{N_{0}}$ of the triangulation $|T_{l}|\rightarrow M$ and let $
\Gamma $ be the ribbon graph associated with the corresponding dual conical
polytope $(|P_{T_{l}}|\rightarrow {M)}$, then the map 
\begin{gather}
\Upsilon :(|P_{T_{l}}|\rightarrow {M)\longrightarrow }((M;N_{0}),\mathcal{C}
);\{\phi (k,t(k)\})  \label{riemsurf} \\
\Gamma \longmapsto \bigcup_{\{\rho ^{0}(j)\}}^{N_{2}(T)}U_{\rho
^{0}(j)}\bigcup_{\{\rho ^{1}(h)\}}^{N_{1}(T)}U_{\rho ^{1}(h)}\bigcup_{\{\rho
^{2}(k)\}}^{N_{0}(T)}(U_{\rho ^{2}(k)},\phi (k,t(k))|_{\rho ^{2}(k)}), 
\notag
\end{gather}
defines the decorated, $N_{0}$-pointed, Riemann surface $((M;N_{0}),
\mathcal{C})$ canonically associated with the conical Regge polytope $
|P_{T_{l}}|\rightarrow {M}$.

\end{proposition}
\begin{figure}[ht]
\begin{center}
\includegraphics[scale=.4]{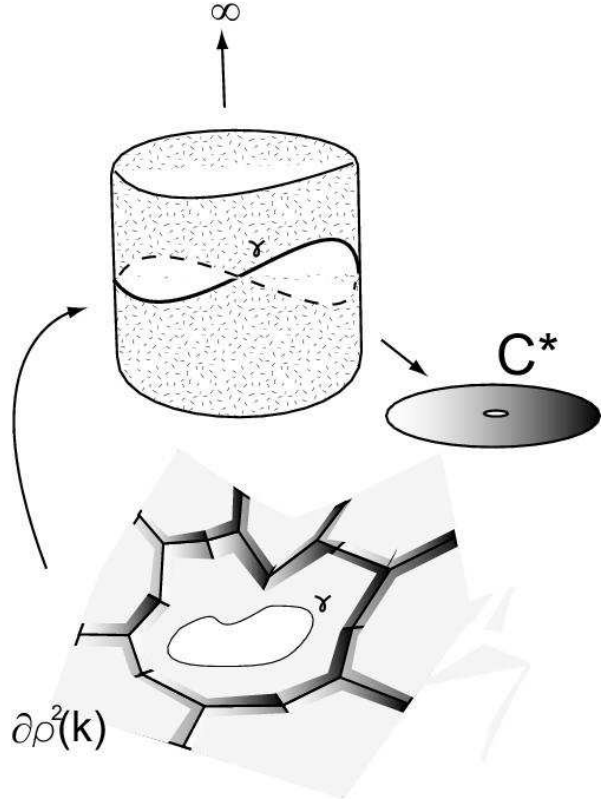}
\caption{The cylindrical metric associated with a ribbon graph.}
\end{center}
\end{figure}

\noindent In other words, through the morphism $\Upsilon $ defined by the
gluing maps (\ref{glue1}) and (\ref{glue2}) one can generate, from a
polytope barycentrically dual to a (generalized) Regge triangulation, a
Riemann surface $((M;N_{0}),\mathcal{C})\in \overline{\mathfrak{M}}_{g},_{N_{0}}$
. Such a surface naturally carries the decoration provided by a choice of
local coordinates $\zeta (k)$ around each puncture, of the corresponding
meromorphic quadratic differential $\phi (k,t(k))|_{\rho ^{2}(k)}$ and of
the associated conical metric $ds_{(k)}^{2}$. It is through such a
decoration that the punctured Riemann surface $((M;N_{0}),\mathcal{C})$
keeps track of the metric geometry of the conical Regge polytope $
|P_{T_{l}}|\rightarrow {M}$ out of which $((M;N_{0}),\mathcal{C})$ has been
generated.  

\bigskip

The above proposition characterizes the spaces of conical Regge polytopes $
\overline{RP}_{g,N_{0}}^{met}$ as a local covering for $\overline{\mathfrak{M}}
_{g},_{N_{0}}$.

\bigskip

\subsection*{\normalsize{\bf 4.2. The Weil-Petersson geometry of Regge Polytopes.}} 
We proceed in a similar vein and describe the Weil-Petersson geometry associated with a
Regge polytope. To this end, let us start by observing that we can represent
finer details of the geometry of $(U_{\rho ^{2}(k)},ds_{(k)}^{2})$ by
opening the cone into its constituent conical sectors. To motivate such a
representation, let $W_{\alpha }(k)$, $\alpha =1,...,q(k)$, be the
barycenters of the edges $\sigma ^{1}(\alpha )\in |T_{l}|\rightarrow {M}$
incident on $\sigma ^{0}(k)$, and intercepting the boundary $\partial (\rho
^{2}(k))$ of the polygonal cell $\rho ^{2}(k)$. Denote by $l(\partial (\rho
^{2}(k)))$ the length of $\partial (\rho ^{2}(k))$, and by $\widehat{L}
_{\alpha }(k)$ the length of the polygonal $\partial (\rho ^{2}(k))$ between
the points $W_{\alpha }(k)$ and $W_{\alpha +1}(k)$ (with $\alpha $ defined $
\func{mod}q(k)$). In the uniformization $\zeta (k)$ of $C|lk(\sigma ^{0}(k))|
$, the points $\{W_{\alpha }(k)\}$ characterize a corresponding set of
points on the circumference $\{\zeta (k)\in \mathbb{C}\;|\;\left| \zeta
(k)\right| =l(\partial (\rho ^{2}(k)))\}$, (for simplicity, we have set $
\zeta _{k}(\sigma ^{0}(k))=0$), and an associated set of $q(k)$ generators $
\{\overline{W_{\alpha }(k)\sigma ^{0}(k)}\}$ on the cone $(U_{\rho
^{2}(k)},ds_{(k)}^{2})$. Such generators mark $q(k)$ conical sectors 
\begin{equation}
S_{\alpha }(k)\doteq \left( c_{\alpha }(k),\frac{l(\partial (\rho ^{2}(k)))}{
\theta (k)},\vartheta _{\alpha }(k)\right) ,
\end{equation}
with base 
\begin{equation}
c_{\alpha }(k)\doteq \left\{ \left| \zeta (k)\right| =l(\partial (\rho
^{2}(k))),\arg W_{\alpha }(k)\leq \arg \zeta (k)\leq \arg W_{\alpha
+1}(k)\right\} ,  \label{consect}
\end{equation}
slant radius $\frac{l(\partial (\rho ^{2}(k)))}{\theta (k)}$, and with
angular opening 
\begin{equation}
\vartheta _{\alpha }(k)\doteq \frac{\widehat{L}_{\alpha }(k)}{l(\partial
(\rho ^{2}(k)))}\theta (k),  \label{angles}
\end{equation}
where $\theta (k)=2\pi -\varepsilon (k)$ is the given conical angle. Since $
\sum_{\alpha =1}^{q(k)}\vartheta _{\alpha }(k)=\theta (k)$, the $\{\vartheta
_{\alpha }(k)\}$ are the representatives, in the uniformization $(U_{\rho
^{2}(k)},ds_{(k)}^{2})$, of the $q(k)$ vertex angles generating the deficit
angle $\varepsilon (k)$ of $C|lk(\sigma ^{0}(k))|$. In particular, we can
formally represent the cone $(U_{\rho ^{2}(k)},ds_{(k)}^{2})$ as 
\begin{equation}
(U_{\rho ^{2}(k)},ds_{(k)}^{2})=\cup _{\alpha =1}^{q(k)}S_{\alpha }(k).
\end{equation}
If we split open the vertex of the cone and of the associated conical
sectors $S_{\alpha }(k)$, then the conical geometry of $(U_{\rho
^{2}(k)},ds_{(k)}^{2})$ can be equivalently described by a cylindrical strip
of height $\frac{l(\partial (\rho ^{2}(k)))}{\theta (k)}$ decorating the
boundary of $\rho ^{2}(k)$. Each sector $S_{\alpha }(k)$ in the cone gives
rise, in such a picture, to a rectangular region in the cylindrical strip.
It is profitable to explicitly represent any such a region, in the complex
plane of the variable $z=x+\sqrt{-1}y$, upside down according to 
\begin{equation}
R_{\vartheta _{\alpha }(k)}(k)\doteq \left\{ z\in \mathbb{C}|\;0\leq x\leq 
\frac{l(\partial (\rho ^{2}(k)))}{\theta (k)},0\leq y\leq \widehat{L}
_{\alpha }(k)\right\} .
\end{equation}

\begin{figure}[ht]
\begin{center}
\includegraphics[scale=.4]{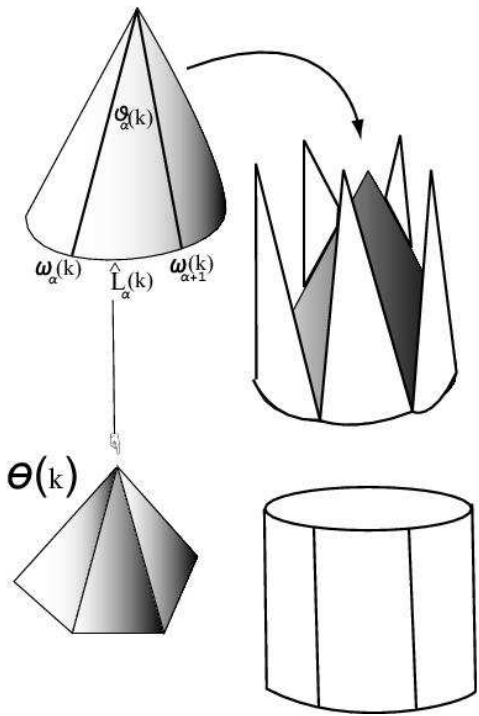}
\caption{The opening of the cone into its constituents conical sectors and the associated cylindrical strip.}
\end{center}
\end{figure}

\noindent We can go a step further, and by means of the conformal transformation 
\begin{equation}
W_{\alpha }(k)=\exp \left[ \frac{2\pi \sqrt{-1}\theta (k)}{l(\partial (\rho
^{2}(k)))}z\right] ,
\end{equation}
we can map the rectangle $R_{\vartheta _{\alpha }(k)}(k)$ into the annulus 
\begin{equation}
\Delta _{\vartheta _{\alpha }(k)}(k)\doteq \{W(k)\in \mathbb{C}|\;|t_{\alpha
}(k)|<\left| W(k)\right| <1\},  \label{annulus}
\end{equation}
where 
\begin{equation}
|t_{\alpha }(k)|\doteq \exp \left[ -2\pi \vartheta _{\alpha }(k)\right] . 
\notag
\end{equation}
Note that 
\begin{equation}
\frac{1}{2\pi }\ln \left( \frac{1}{|t_{\alpha }(k)|}\right) =\vartheta
_{\alpha }(k),  \label{modulann}
\end{equation}
is the modulus of $\Delta _{\vartheta _{\alpha }(k)}(k)$.

\begin{figure}[ht]
\begin{center}
\includegraphics[scale=.4]{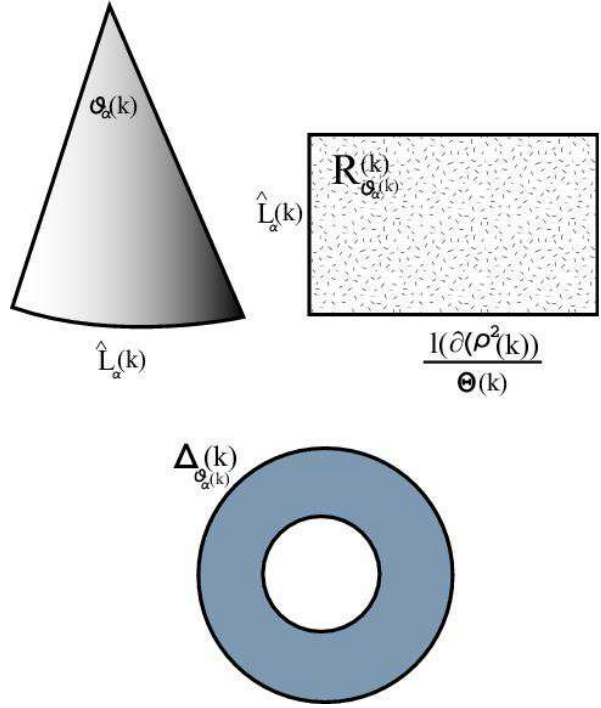}
\caption{The mapping of a conical sector in a strip and into an anular region.}
\end{center}
\end{figure}

Such a remark further motivates the analysis of the geometry of (random)
Regge triangulations from the point of view of moduli theory. In this
connection note that for a given set of perimeters $\{l(\partial (\rho
^{2}(k)))\}_{k=1}^{N_{0}}$ and deficit angles $\{\varepsilon
(k)\}_{k=1}^{N_{0}}$ there are $(N_{1}(T)-N_{0}(T))$ free angles $\vartheta
_{\alpha }(k)$. This follows by observing that, for given $l(\partial (\rho
^{2}(k)))$ and $\varepsilon (k)$, the angles $\vartheta _{\alpha }(k)$ are
characterized (see (\ref{angles})) by the $\widehat{L}_{\alpha }(k)$. These
latter are in a natural correspondence with the $N_{1}$ edges of $
|P_{T_{l}}|\rightarrow {M}$, and among them we have the $N_{0}$ constraint $
\sum_{\alpha }^{q(k)}\widehat{L}_{\alpha }(k)=l(\partial (\rho ^{2}(k)))$.
From $N_{0}(T)-N_{1}(T)+N_{2}(T)=2-2g$, and the relation $
2N_{1}(T)=3N_{2}(T) $ associated with the trivalency, we get $
N_{1}(T)-N_{0}(T)=2N_{0}(T)+6g-6$, which exactly corresponds to the real
dimension of the moduli space $\mathfrak{M}_{g},_{N_{0}}$ of genus $g$
Riemann surfaces $((M;N_{0}),\mathcal{C})$ with $N_{0}$ punctures.

According to the above procedure, we can open the cones $(U_{\rho
^{2}(k)},ds_{(k)}^{2})=\cup _{\alpha =1}^{q(k)}S_{\alpha }(k)$ and map each
conical sector $S_{\alpha }(k)$ onto a corresponding annulus 
\begin{equation}
\Delta _{\vartheta _{\alpha }(k)}(k)\doteq \{W(k)\in \mathbb{C}\;|\;e^{-2\pi
\vartheta _{\alpha }(k)}<\left| W(k)\right| <1\}.  \label{ann2}
\end{equation}
In such a setting, a natural question to discuss concerns how a deformation
of the conical sectors $S_{\alpha }(k)$, at fixed deficit angle $\varepsilon
(k)$ (and perimeter $l(\partial (\rho ^{2}(k)))$), affects the conical
geometry of $(U_{\rho ^{2}(k)},ds_{(k)}^{2})$, and how this deformation
propagates to the underlying Regge polytope $|P_{T_{l}}|\rightarrow {M}$.
The question is analogous to the study of the non-trivial deformations of
constant curvature metrics on a surface. To this end, we adapt to our
purposes a standard procedure by considering the following deformation of $
S_{\alpha }(k)$ 
\begin{equation}
\vartheta _{\alpha }(k)\longmapsto \vartheta _{\alpha }(k)^{\prime }\doteq
(s+1)\vartheta _{\alpha }(k),\;s\in \mathbb{R},
\end{equation}
and discuss its effect on the map $\Upsilon $ near the identity $s=0$. In
the annulus description of the geometry of the corresponding sector $
S_{\alpha }(k)$, such a deformation is realized by the quasi-conformal map 
\begin{equation}
W(k)\longmapsto f(W(k))=W(k)\left| W(k)\right| ^{s},
\end{equation}
which indeed maps the annulus (\ref{ann2}) into the annulus 
\begin{equation}
\{W(k)\in \mathbb{C}\;|\;e^{-2\pi (s+1)\vartheta _{\alpha }(k)}<\left|
W(k)\right| <1\},
\end{equation}
corresponding to a conical sector of angle $\vartheta _{\alpha }(k)^{\prime
}\doteq (s+1)\vartheta _{\alpha }(k)$. The study of the transformation $
f(W(k))$ is well known in the modular theory of the annulus (see \cite
{wolpert}), and goes as follows. The Beltrami differential associated with $
f(\zeta (k))$ at $s=0$, is given by 
\begin{equation}
\frac{\partial }{\partial \overline{W(k)}}\left[ \left. \frac{\partial }{
\partial s}f(W(k))\right| _{s=0}\right] =\frac{W(k)}{2\overline{W(k)}}.
\end{equation}
The corresponding Beltrami differential on (\ref{ann2}) representing the
infinitesimal deformation of $S_{\alpha }(k)$ in the direction $\partial
/\partial (e^{-2\pi \vartheta _{\alpha }(k)})$ is provided by 
\begin{eqnarray}
\mu _{\alpha }(k) &\doteq &\left( \frac{\partial e^{-2\pi (s+1)\vartheta
_{\alpha }(k)}}{\partial s}\right) _{s=0}^{-1}\frac{\partial }{\partial 
\overline{W(k)}}\left[ \left. \frac{\partial }{\partial s}f(W(k))\right|
_{s=0}\right] = \\
&=&-\left( \frac{1}{2\pi \vartheta _{\alpha }(k)}\right) e^{2\pi \vartheta
_{\alpha }(k)}\frac{W(k)}{2\overline{W(k)}}.  \notag
\end{eqnarray}

\begin{figure}[ht]
\begin{center}
\includegraphics[scale=.4]{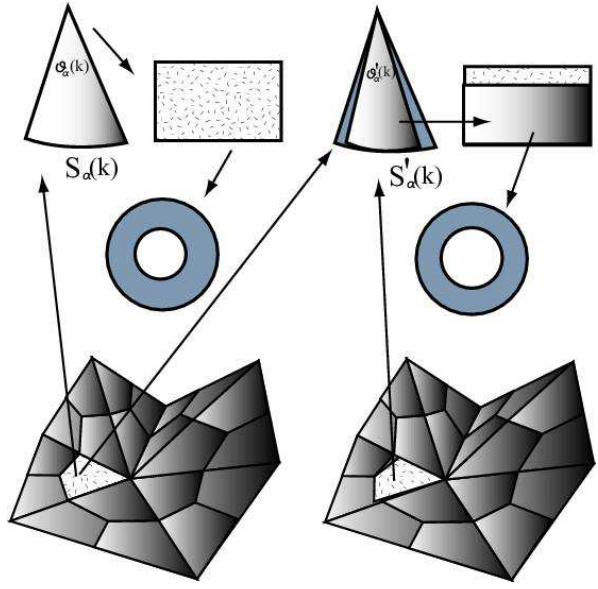}
\caption{The modular deformation of a conical sector in a Regge polytope.}
\end{center}
\end{figure}

\noindent Recall that a Beltrami differential on the annulus is naturally
paired with a quadratic differential $\widehat{\phi _{\alpha }(k)}\doteq
C_{k}\frac{dW(k)\otimes dW(k)}{W^{2}(k)}$, $C_{k}$ being a real constant,
via the $L^{2} $ inner product 
\begin{equation}
\left\langle \mu _{\alpha }(k),\widehat{\phi _{\alpha }(k)}\right\rangle
_{\Delta _{\vartheta _{\alpha }(k)}(k)}\doteq \frac{\sqrt{-1}}{2}
\int_{\Delta _{\vartheta _{\alpha }(k)}(k)}\mu (k)\widehat{\phi _{\alpha }(k)
}dW(k)\wedge d\overline{W(k)}.
\end{equation}
By requiring that 
\begin{equation}
\left\langle \mu _{\alpha }(k),\widehat{\phi _{\alpha }(k)}\right\rangle
_{\Delta _{\vartheta _{\alpha }(k)}(k)}=1
\end{equation}
one finds the constant $C_{k}$ characterizing the quadratic differential $
\widehat{\phi _{\alpha }(k)}$, dual to $\mu _{\alpha }(k)$, and describing
the infinitesimal deformation of $S_{\alpha }(k)$ in the direction $
d(e^{-2\pi \vartheta _{\alpha }(k)})$, (\emph{i.e.}, a cotangent vector to $
\mathfrak{M}_{g},_{N_{0}}$ at $((M;N_{0}),\mathcal{C});\{ds_{(k)}^{2}\})$).
A direct computation provides 
\begin{equation}
\widehat{\phi _{\alpha }(k)}=-\frac{e^{-2\pi \vartheta _{\alpha }(k)}}{\pi }
\frac{dW(k)\otimes dW(k)}{W^{2}(k)}.
\end{equation}
Note that whereas $\phi (k)_{\rho ^{2}(k)}$ may be thought of as a cotangent
vector to one of the $\mathbb{R}^{N_{0}}$ fiber of $\mathfrak{M}
_{g},_{N_{0}}\times \mathbb{R}_{+}^{N}$, the quadratic differential $
\widehat{\phi _{\alpha }(k)}$projects down to a deformation in the base $
\mathfrak{M}_{g},_{N_{0}}$, and as such is a much more interesting object
than $\phi (k)_{\rho ^{2}(k)}$. The modular nature of $\widehat{\phi
_{\alpha }(k)}$ comes clearly to the fore if, along the lines of section 3.3, we compute its associated
Weil-Petersson norm according to 
\begin{equation}
\left\| \widehat{\phi _{\alpha }(k)}\right\| _{W-P}=\int_{\Delta _{\vartheta
_{\alpha }(k)}(k)}\frac{|\widehat{\phi _{\alpha }(k)}|^{2}}{g_{hyp}},
\end{equation}
where $g_{hyp}$ denotes the hyperbolic metric 
\begin{equation}
g_{hyp}\doteq \left( \frac{\pi ^{2}}{\ln ^{2}|t_{\alpha }(k)|}\right) \frac{
dW(k)d\overline{W(k)}}{|W(k)|^{2}\sin ^{2}\left( \pi \frac{\ln |W(k)|}{\ln
|t_{\alpha }(k)|}\right) }
\end{equation}
on the annulus $\Delta _{\vartheta _{\alpha }(k)}(k)$, and where $|t_{\alpha
}(k)|\doteq e^{-2\pi \vartheta _{\alpha }(k)}$. A direct computation
provides \cite{wolpert} 
\begin{equation}
\left\| \widehat{\phi _{\alpha }(k)}\right\| _{W-P}=\left| t_{\alpha
}(k)\right| ^{2}\left( \frac{1}{\pi }\ln \frac{1}{|t_{\alpha }(k)|}\right)
^{3}=8\vartheta _{\alpha }^{3}(k)e^{-4\pi \vartheta _{\alpha }(k)}.
\label{WPnomr}
\end{equation}
According to proposition \ref{gluing}, the decorated Riemann surface 
\begin{equation*}
(((M;N_{0}),\mathcal{C});\{ds_{(k)}^{2}\})
\end{equation*}
is generated by glueing the oriented domains $(U_{\rho
^{2}(k)},ds_{(k)}^{2}) $ along the ribbon graph $\Gamma $ associated with
the Regge polytope $|P_{T_{l}}|\rightarrow {M}$. We can indipendently and
holomorphically open the distinct cones $\{(U_{\rho ^{2}(k)},ds_{(k)}^{2})\}$
on corresponding sectors $\{S_{\alpha }(k)\}$, then it follows that to any
such $(((M;N_{0}),\mathcal{C});\{ds_{(k)}^{2}\})$ we can associate a well
defined Weil-Petersson metric which can be immediately read off from (\ref
{WPnomr}). Since there are $6g-6+2N_{0}$ independent conical sectors $
S_{\alpha }(k)$ on the Riemann surface $(((M;N_{0}),\mathcal{C}
);\{ds_{(k)}^{2}\})$ corresponding to the Regge polytope $
|P_{T_{l}}|\rightarrow {M}$, (at fixed deficit angles $\{\varepsilon (k)\}$
and perimeters $\{l(\partial (\rho ^{2}(k)))\}$), we can find a
corresponding sequence of angles $\{\vartheta _{\alpha }(k)\}$ and relabel
them as $\{\vartheta _{H}\}_{H=1}^{6g-6+2N_{0}}$. If we define 
\begin{equation}
\tau _{H}\doteq e^{-2\pi \vartheta _{H}}
\end{equation}
then we can immediately write the Weil-Petersson metric $ds_{W-P}^{2}$
associated with the Regge polytope $|P_{T_{l}}|\rightarrow {M}$ 
\begin{eqnarray}
ds_{W-P}^{2}(|P_{T_{l}}|) &=&\sum_{H=1}^{6g-6+2N_{0}}\frac{2\pi ^{3}d\tau
_{H}^{2}}{\tau _{H}^{2}\left( \ln \frac{1}{\tau _{H}}\right) ^{3}}=
\label{WPmetr1} \\
&=&\pi ^{2}\sum_{H=1}^{6g-6+2N_{0}}\frac{d\vartheta _{H}^{2}}{\vartheta
_{H}^{3}}.  \notag
\end{eqnarray}
If we respectively denote by $\theta (H)$ and $l(H)$ the (fixed) conical
angle $\theta (k)$ and the perimeter $l(\partial (\rho ^{2}(k)))$
corresponding to the modular variable $\vartheta _{H}$, (note that the same $
\theta (H)$ and $l(H)$ correspond to the distinct $\{\vartheta _{H}\}$ which
are incident on the same vertex), then according to (\ref{angles}) we can
rewrite (\ref{WPmetr1}) as 
\begin{equation}
ds_{W-P}^{2}(|P_{T_{l}}|)=\sum_{H=1}^{6g-6+2N_{0}}\frac{\pi ^{2}}{\theta (H)}
\left( \frac{\widehat{L}_{H}}{l(H)}\right) ^{-3}d\left( \frac{\widehat{L}_{H}
}{l(H)}\right) \otimes d\left( \frac{\widehat{L}_{H}}{l(H)}\right) ,
\label{WPmetr2}
\end{equation}
where $\widehat{L}_{H}$ is the length variable $\widehat{L}_{\alpha }(k)$
associated with $\vartheta _{H}$.

Note that (\ref{WPmetr1}) and (\ref{WPmetr2}) show a singular behavior when $
\vartheta _{H}\rightarrow 0$ (or equivalently when $\widehat{L}
_{H}\rightarrow 0$), such a behavior occurs for instance when the Regge
triangulation (and the corresponding polytope) degenerates and exhibits
vertices where thinner and thinner triangles are incident. As recalled in
paragraph 2.3, this is usually considered a pathology of
Regge calculus. However, in view of the modular correspondence we have
described in this paper, it simply corresponds to the well-known
incompleteness (with respect of the complex structure of the moduli space)
of the Weil-Petersson metric on $\mathfrak{M}_{g},_{N_{0}}$ as we approach
the boundary (the compactifying divisor) of $\mathfrak{M}_{g},_{N_{0}}$ in $
\overline{\mathfrak{M}}_{g},_{N_{0}}$ . In such a sense, as already remarked
in the introductory remarks, such a pathology of Regge triangulations has a
modular meaning and is not accidental.

\subsection*{\normalsize{\bf 4.3. The W-P volume on the space of Regge polytopes.}}
With the metric $ds_{W-P}^{2}(|P_{T_{l}}|)$ we can associate a well defined
volume form 
\begin{gather}
\Omega _{W-P}(|P_{T_{l}}|)= \\
=\left[ \det \left( \frac{\pi ^{2}}{\theta (H)}\left( \frac{\widehat{L}_{H}}{
l(H)}\right) ^{-3}\right) \right] ^{\frac{1}{2}}d\left( \frac{\widehat{L}_{1}
}{l(1)}\right) \bigwedge ...\bigwedge d\left( \frac{\widehat{L}_{6g-6+2N_{0}}
}{l(6g-6+2N_{0})}\right) ,  \notag
\end{gather}
which can be rewritten as a power of product of $2$-forms of the type $
d\left( \frac{\widehat{L}_{H}}{l(H)}\right) \wedge d\left( \frac{\widehat{L}
_{H+1}}{l(H+1)}\right) $, (this being connected with the K\"{a}hler form
associated with (\ref{WPmetr2})). Such forms are directly related with the
Chern classes (\ref{chern}) of the line bundles $\mathcal{CL}_{k}$ and play
a distinguished role in the Kontsevich-Witten model. The possibility of
expressing the Weil-Petersson K\"{a}hler form in terms of the Chern classes (\ref{chern}) 
is a deep and basic fact of the geometry of $\overline{
\mathfrak{M}}_{g},_{N_{0}}$, and it is a pleasant feature of the model
discussed here that such a connection can be motivated by rather elementary
considerations.

The Weil-Petersson volume form $\Omega _{W-P}(|P_{T_{l}}|)$ allows us to
integrate over the space 
\begin{equation}
RP_{g,N_{0}}^{met}(\{\varepsilon (H)\},\{l(H)\})
\end{equation}
of distinct Regge polytopes $|P_{T_{l}}|\rightarrow {M}$ with given deficit
angles $\{\varepsilon (k)\}$ and perimeters $\{l(\partial (\rho ^{2}(k)))\}$
. To put such an integration in a proper perspective and give it a
suggestive physical meaning we shall explicitly consider the set of Regge
polytopes 
\begin{equation}  \label{RPg}
RP_{g,N_{0}}^{met}(\{q(H)\})\doteq \left\{ |P_{T_{l}}|\rightarrow {M|\;}
\varepsilon (H)=2\pi -\frac{\pi }{3}q(H);\;l(H)=\frac{\sqrt{3}}{3}
aq(H)\right\} ,
\end{equation}
which contain the equilateral Regge polytopes dual to dynamical
triangulations. According to proposition \ref{gluing}, $RP_{g,N_{0}}^{met}(
\{q(H)\})$ is a combinatorial description of $\mathfrak{M}_{g},_{N_{0}}$,
and the volume form $\Omega _{W-P}(|P_{T_{l}}|)$ can be considered as the
pull-back under the morphism $\Upsilon $, (see (\ref{riemsurf})) of the
Weil-Petersson volume form $\omega _{WP}^{3g-3+N_{0}}/(3g-3+N_{0})!$ on $
\mathfrak{M}_{g},_{N_{0}}$.

Stated differently, the integration of $\Omega _{W-P}(|P_{T_{l}}|)$ over the
space (\ref{RPg}) is equivalent to the Weil-Petersson volume of $\mathfrak{M}
_{g},_{N_{0}}$. It is well-known that the Weil-Petersson form $\omega _{WP}$
extends (as a (1,1) current) to the compactification $\overline{\mathfrak{M}}
_{g},_{N_{0}}$, and, if we denote by $\overline{RP}_{g,N_{0}}^{met}(\{q(H)
\}) $ the compactified orbifold associated with $RP_{g,N_{0}}^{met}(\{q(H)\})
$, then we can write 
\begin{equation*}
\frac{1}{N_{0}!}\int_{\overline{RP}_{g,N_{0}}^{met}(\{q(H)\})}\Omega
_{W-P}(|P_{T_{l}}|)=
\end{equation*}
\begin{equation}  \label{integ}
=\frac{1}{N_{0}!}\int_{\overline{\mathfrak{M}}_{g},_{N_{0}}}\frac{\omega
_{WP}^{3g-3+N_{0}}}{(3g-3+N_{0})!}=Vol\left( \overline{\mathfrak{M}}
_{g},_{N_{0}}\right)
\end{equation}
where we have divided by $N_{0}(T)!$ in order to factor out the labelling of
the $N_{0}(T)$ punctures. Since $\overline{RP}_{g,N_{0}}^{met}(\{q(H)\})$ is
a (smooth) stratified orbifold acted upon by the automorphism group $
Aut_{\partial }(P_{T_{a}})$, we can explicitly write the left side of (\ref
{integ}) as an orbifold integration over the distinct orbicells $\Omega
_{T_{a}}$ (\ref{strata}) in which $\overline{RP}_{g,N_{0}}^{met}(\{q(H)\})$
is stratified

\begin{gather}
\frac{1}{N_{0}!}\sum_{T\in \mathcal{DT}[\{q(i)\}_{i=1}^{N_{0}}]}\frac{1}{
|Aut_{\partial }(P_{T_{a}})|}\int_{_{\Omega
_{T_{a}}(\{q(k)\}_{k=1}^{N_{0}})}}\Omega _{W-P}(|P_{T_{l}}|)=  \label{orbin}
\\
\notag \\
=VOL\left( \overline{\mathfrak{M}}_{g},_{N_{0}}\right) ,  \notag
\end{gather}
(the orbifold integration is defined in \cite{penner}, Th. 3.2.1), where the
summation is over all distinct dynamical triangulations with given unlabeled
curvature assignments weighted by the order $|Aut_{\partial }(P_{T})|$ of
the automorphisms group of the corresponding dual polytope. The relation 
(\ref{orbin}) provides a non-trivial connection between dynamical
triangulations (labelling the strata of $\overline{RP}_{g,N_{0}}^{met}(
\{q(H)\})$ (or, equivalently, of $\overline{\mathfrak{M}}_{g},_{N_{0}}$),
and the fixed connectivity Regge triangulations in each strata $\Omega
_{T_{a}}$. Some aspects of this relation have already been discussed by us
elsewhere, \cite{carfora2}. Here we will exploit (\ref{orbin}) in order to
directly relate $VOL\left( \overline{\mathfrak{M}}_{g},_{N_{0}}\right) $ to
the partition function of 2D simplicial quantum gravity. To this end let us
sum both members of (\ref{orbin}) over the set of all possible curvature
assignments $\{q(H)\}$ on the $N_{0}$ unlabelled vertices of the
triangulations, and note that 
\begin{multline}
Card\left[ \mathcal{DT}(N_{0})\right] \doteq \sum_{\mathcal{DT}(N_{0})}\frac{
1}{|Aut_{\partial }(P_{T_{a}})|}= \\
=\frac{1}{N_{0}!}\sum_{\{q(H)\}_{H=1}^{N_{0}}}\sum_{T\in \mathcal{DT}
[\{q(H)\}_{H=1}^{N_{0}}]}\frac{1}{|Aut_{\partial }(P_{T_{a}})|}  \notag
\end{multline}
provides the number of distinct (generalized) dynamical triangulations with $
N_{0}$ unlabelled vertices. Since $VOL\left( \overline{\mathfrak{M}}
_{g},_{N_{0}}\right) $ does not depend of the curvature assignments $
\{q(H)\} $, from (\ref{orbin}) we get 
\begin{gather}
\frac{1}{N_{0}!}\sum_{\{q(H)\}_{H=1}^{N_{0}}}\sum_{T\in \mathcal{DT}
[\{q(i)\}_{i=1}^{N_{0}}]}\frac{1}{|Aut_{\partial }(P_{T_{a}})|}
\int_{_{\Omega _{T_{a}}(\{q(k)\}_{k=1}^{N_{0}})}}\Omega _{W-P}(|P_{T_{l}}|)=
\\
=\sum_{\mathcal{DT}(N_{0})}\frac{1}{|Aut_{\partial }(P_{T_{a}})|}
\int_{_{\Omega _{T_{a}}(\{q(k)\}_{k=1}^{N_{0}})}}\Omega _{W-P}(|P_{T_{l}}|)=
\notag \\
=\left( Card\{q(H)\}\right) VOL\left( \overline{\mathfrak{M}}
_{g},_{N_{0}}\right) ,  \notag
\end{gather}
where $Card\{q(H)\}$ denotes the number of possible curvature assignments on
the $N_{0}$ unlabelled vertices of the triangulations. Dividing both members
by $\left(Card\{q(H)\}\right) $, we eventually get the relation 
\begin{equation}
\sum_{\mathcal{DT}(N_{0})}\frac{1}{|Aut_{\partial }(P_{T_{a}})|}
\int_{_{\Omega _{T_{a}}(\{q(k)\}_{k=1}^{N_{0}})}}\frac{\Omega
_{W-P}(|P_{T_{l}}|)}{Card\{q(H)\}}=VOL\left( \overline{\mathfrak{M}}
_{g},_{N_{0}}\right) ,  \label{partfunc}
\end{equation}
(the number $Card\{q(H)\}$ has been shifted under the integral sign for
typographical convenience). We have the following

\begin{lemma}
For $N_{0}$ sufficiently large there exist a positive constant $K(g)$
independent from $N_{0}$ but possibly dependent on the genus $g$, and a
positive constant $\kappa $ independent both from $N_{0}$ and $g$ such that 
\begin{equation}
\int_{_{\Omega _{T_{a}}(\{q(k)\}_{k=1}^{N_{0}})}}\frac{\Omega
_{W-P}(|P_{T_{l}}|)}{Card\{q(H)\}}\simeq K(g)e^{-\kappa N_{0}}.
\end{equation}
\end{lemma}

\bigskip

In order to prove this result let us start by recalling that the large $
N_{0}(T)$ asymptotics of the triangulation counting $Card\left[ \mathcal{DT}
[N_{0}]\right] $ can be obtained from purely combinatorial (and matrix
theory) arguments \cite{mulase}, \cite{brezin} to the effect that 
\begin{equation}
Card\left[ \mathcal{DT}[N_{0}]\right] \sim \frac{16c_{g}}{3\sqrt{2\pi }}
\cdot e^{\mu _{0}N_{0}(T)}N_{0}(T)^{\frac{5g-7}{2}}\left( 1+O(\frac{1}{N_{0}}
)\right) ,  \label{oldentr}
\end{equation}
where $c_{g}$ is a numerical constant depending only on the genus $g$, and $
e^{\mu _{0}}=(108\sqrt{3})$ is a (non-universal) parameter depending on the
set of triangulations considered (here the generalized triangulations,
barycentrically dual to trivalent graphs; in the case of regular
triangulations in place of $108\sqrt{3}$ we would get $e^{\mu _{0}}=(\frac{
4^{4}}{3^{3}})$). Thus, if we denote by 
\begin{gather}
\left\langle \int \frac{\Omega _{W-P}(|P_{T_{l}}|)}{Card\{q(H)\}}
\right\rangle _{\mathcal{DT}[N_{0}]}\doteq \\
=\frac{1}{Card\left[ \mathcal{DT}[N_{0}]\right] }\sum_{\mathcal{DT}(N_{0})}
\frac{1}{|Aut_{\partial }(P_{T_{a}})|}\int_{_{\Omega
_{T_{a}}(\{q(k)\}_{k=1}^{N_{0}})}}\frac{\Omega _{W-P}(|P_{T_{l}}|)}{
Card\{q(H)\}}  \notag
\end{gather}
the average value of $\int $\ $\Omega _{W-P}(|P_{T_{l}}|)/Card\{q(H)\}$ over
the set $\mathcal{DT}[N_{0}]$, (dropping the integration range $\Omega
_{T_{a}}(\{q(k)\}_{k=1}^{N_{0}})$ for notational ease), then we can write
the large $N_{0}$ asymptotics of the left side member of (\ref{orbin}) as 
\begin{gather}
\sum_{\mathcal{DT}(N_{0})}\frac{1}{|Aut_{\partial }(P_{T_{a}})|}
\int_{_{\Omega _{T_{a}}(\{q(k)\}_{k=1}^{N_{0}})}}\frac{\Omega
_{W-P}(|P_{T_{l}}|)}{Card\{q(H)\}}\simeq  \label{aver} \\
\simeq \frac{16c_{g}}{3\sqrt{2\pi }}\left\langle \int \frac{\Omega
_{W-P}(|P_{T_{l}}|)}{Card\{q(H)\}}\right\rangle _{\mathcal{DT}[N_{0}]}e^{\mu
_{0}N_{0}(T)}N_{0}(T)^{\frac{5g-7}{2}}\left( 1+O(\frac{1}{N_{0}})\right) . 
\notag
\end{gather}
On the other hand, from the Manin-Zograf asymptotic analysis of $VOL\left( 
\overline{\mathfrak{M}}_{g},_{N_{0}}\right) $ for fixed genus $g$ and large $
N_{0}$, we have\cite{manin}\cite{zograf} 
\begin{gather}
VOL\left( \overline{\mathfrak{M}}_{g},_{N_{0}}\right) =  \label{manzog} \\
=\pi ^{2(3g-3+N_{0})}(N_{0}+1)^{\frac{5g-7}{2}}C^{-N_{0}}\left(
B_{g}+\sum_{k=1}^{\infty }\frac{B_{g,k}}{(N_{0}+1)^{k}}\right) ,  \notag
\end{gather}
where $C=-\frac{1}{2}j_{0}\frac{d}{dz}J_{0}(z)|_{z=j_{0}}$, ($J_{0}(z)$ the

Bessel function, $j_{0}$ its first positive zero); (note that $C\simeq
0.625....$). The genus dependent parameters $B_{g}$ are explicitly given 
\cite{zograf} by 
\begin{equation}
\left\{ 
\begin{tabular}{ll}
$B_{0}=\frac{1}{A^{1/2}\Gamma (-\frac{1}{2})C^{1/2}},$ & $B_{1}=\frac{1}{48}
, $ \\ 
$B_{g}=\frac{A^{\frac{g-1}{2}}}{2^{2g-2}(3g-3)!\Gamma (\frac{5g-5}{2})C^{
\frac{5g-5}{2}}}\left\langle \tau _{2}^{3g-3}\right\rangle ,$ & $g\geq 2$
\end{tabular}
\right.
\end{equation}
where $A\doteq -j_{0}^{-1}J_{0}^{\prime }(j_{0})$, and $\left\langle \tau
_{2}^{3g-3}\right\rangle $ is a Kontsevich-Witten intersection number \cite
{kontsevich}\cite{witten}, (the coefficients $B_{g,k}$ can be computed
similarly-see \cite{zograf} for details). By Inserting (\ref{manzog}) in the
left hand side of (\ref{orbin}) and by taking into account (\ref{aver}) we
eventually get 
\begin{gather}
\frac{16c_{g}}{3\sqrt{2\pi }}\left\langle \int \frac{\Omega
_{W-P}(|P_{T_{l}}|)}{Card\{q(H)\}}\right\rangle _{\mathcal{DT}[N_{0}]}e^{\mu
_{0}N_{0}(T)}N_{0}(T)^{\frac{5g-7}{2}}\left( 1+O(\frac{1}{N_{0}})\right) = \\
=\pi ^{2(3g-3+N_{0})}(N_{0}+1)^{\frac{5g-7}{2}}C^{-N_{0}}\left(
B_{g}+\sum_{k=1}^{\infty }\frac{B_{g,k}}{(N_{0}+1)^{k}}\right) ,  \notag
\end{gather}
which by direct comparison provides 
\begin{equation}
\left\langle \int \frac{\Omega _{W-P}(|P_{T_{l}}|)}{Card\{q(H)\}}
\right\rangle _{\mathcal{DT}[N_{0}]}\simeq \frac{3\sqrt{2\pi }B_{g}\pi
^{2(3g-3+N_{0})}}{16c_{g}}e^{(|\ln C|-\mu _{0})N_{0}}.
\end{equation}
This proves the lemma with 
\begin{equation}
K(g)=\frac{3\sqrt{2\pi }B_{g}\pi ^{2(3g-3)}}{16c_{g}},
\end{equation}
and 
\begin{equation}
\kappa =\mu _{0}-|\ln C|-2\ln\pi\approx 2.472.  \label{kappa}
\end{equation}
Thus, according to the above lemma, we can rewrite the left member of (\ref
{partfunc}) as 
\begin{gather}
VOL\left( \overline{\mathfrak{M}}_{g},_{N_{0}}\right) =\sum_{\mathcal{DT}
(N_{0})}\frac{1}{|Aut_{\partial }(P_{T_{a}})|}\int_{_{\Omega
_{T_{a}}(\{q(k)\}_{k=1}^{N_{0}})}}\frac{\Omega _{W-P}(|P_{T_{l}}|)}{
Card\{q(H)\}}=  \label{DTpart} \\
=K(g)\sum_{\mathcal{DT}(N_{0})}\frac{1}{|Aut_{\partial }(P_{T_{a}})|}
e^{-\kappa N_{0}},  \notag
\end{gather}
which has the structure of the canonical partition function for dynamical
triangulation theory \cite{ambjorn}. Roughly speaking, we may interpret (\ref
{DTpart}) by saying that dynamical triangulations count the (orbi)cells
which contribute to the volume (each cell containing the Regge
triangulations whose adjacency matrix is that one of the dynamical
triangulation which label the cell), whereas integration over the cells
weights the fluctuations due to all Regge triangulations which, in a sense,
represent the deformational degrees of freedom of the given dynamical
triangulation labelling the cell.

\subsection*{\normalsize{\bf 4.4. The Hodge-Deligne decomposition.}}
Further properties of triangulated surfaces have their origin in the Hodge geometry
of the corresponding Riemann surface. To this end, let us consider the
collection of \ $N_{0}$ meromorphic 1-forms associated with the quadratic
differentials $\{\phi (k)\}$, 
\begin{equation}
\left( \sqrt{\phi (1)},...,\sqrt{\phi (N_{0})}\right) \in \Omega
^{1}(\{p_{j}\}),
\end{equation}
where $\{p_{j}\}$ is the divisor of $\sqrt{\phi }$, and $\Omega
^{1}(\{p_{j}\})$ denotes the space of meromorphic 1-forms $\vartheta $ with
divisor $(D_{\vartheta })\leq \{p_{j}\}$. In each uniformization $\zeta (k)$
\ we\ have \ 
\begin{equation}
\sqrt{\phi (k)}\doteq \frac{\sqrt{-1}}{2\pi }\frac{l(\partial (\rho ^{2}(k)))
}{\zeta (k)}d\zeta (k)\underset{DT}{=}\frac{\sqrt{-1}}{2\pi }\frac{\left( 
\frac{\sqrt{3}}{3}a\right) q(k)}{\zeta (k)}d\zeta (k),
\end{equation}
where the subscript $DT$ refers to dynamical triangulations.
According to (\ref{length}) the residues of $\{\sqrt{\phi (k)}\}$ provide
the perimeters $\{l(\partial (\rho ^{2}(k)))\}$. However, the role of $\{
\sqrt{\phi (k)}\}$ is more properly seen in connection with the
introduction, on the cohomology group $H^{1}(M,N_{0};\mathbb{C})=H^{1}(M_{k};
\mathbb{R})\tbigotimes_{\mathbb{R}}\mathbb{C}$, of a Hodge structure
analogous to the classical Hodge decomposition of $H^{h}(M;\mathbb{C})$
generated by the spaces $\mathcal{H}^{r,h-r}$ of harmonic $h$-forms on $M$
of type $(r,h-r)$. Such a decomposition does not hold, as it stands, for
punctured surfaces (since $H^{1}(M,N_{0};\mathbb{C})$ can be
odd-dimensional), but it can be replaced by a mixed Deligne-Hodge
decomposition such that
\begin{equation}
H^{1}(M,N_{0};\mathbb{C})\doteq \tbigoplus_{p,r}I^{p,r},
\label{Deligne1}
\end{equation}
where 
\begin{equation}
\begin{tabular}{ccccc}
$I^{1,1}=\Omega ^{1}(\{p_{j}\})\cap \overline{\Omega ^{1}(\{p_{j}\})},$ &  & 
$I^{1,0}=H^{1,0}(M),$ &  & $I^{0,1}=H^{0,1}(M)$
\end{tabular}
\label{Deligne2}
\end{equation}
with $\dim _{C}I^{1,1}=N_{0}-1$, (this dimensionality is strictly related
with the constraint (\ref{rediv})), and where $H^{1,0}(M)\simeq
H^{0}(M,\Omega ^{1})$, ($H^{0,1}(M)\simeq \overline{H^{0}(M,\Omega ^{1})}$
), denotes the space of holomorphic 1-forms on $M$. More explicitly, the
subspace $I^{1,1}$ can be characterized as 
\begin{equation}
I^{1,1}\cap H^{1}(M,N_{0};\mathbb{R})=\left\{ \sqrt{-1}\frac{\partial h}{
\partial z}dz\,\;|\;h\in \mathcal{H}^{0}(M,N_{0})\right\} ,
\end{equation}
where $\mathcal{H}^{0}(M,N_{0})$ denotes the space of real-valued harmonic
functions on $M$ which have at worst logarithmic singularities along the $
N_{0}$ punctures \cite{pearlstein}. It follows that the forms $\{\sqrt{\phi (k)}\}_{k=1}^{N_{0}}$  are
in $I^{1,1}$ (at the level of cohomology, \emph{i.e.} up to exact forms),
for we can write 
\begin{gather}
\sqrt{\phi (k)}-\frac{\sqrt{-1}}{2\pi }l(\partial (\rho ^{2}(k)))d\ln \left|
\zeta (k)\right| = \\
=\frac{\sqrt{-1}}{4\pi }l(\partial (\rho ^{2}(k)))\left( \frac{d\zeta (k)}{
\zeta (k)}-\frac{d\overline{\zeta }(k)}{\overline{\zeta }(k)}\right) \in
\Omega ^{1}(\{p_{j}\})\cap \overline{\Omega ^{1}(\{p_{j}\})}.  \notag
\end{gather}
Therefore, if we fix a puncture, say $p_{N_{0}}$, and
consider the $N_{0}-1$ cohomology classes 
\begin{equation}
\vartheta _{k}\doteq \frac{1}{2\pi \sqrt{-1}}\left[ \frac{d\zeta (N_{0})}{
\zeta (N_{0})}-\frac{d\zeta (k)}{\zeta (k)}-d\ln \frac{\left| \zeta
(N_{0})\right| }{\left| \zeta (k)\right| }\right] ,  \label{basis}
\end{equation}
then $\{\vartheta _{k}\}_{k=1}^{N_{0}-1}$ defines a basis for $I^{1,1}$ in
terms of the complex coordinates $\{\zeta (k)\}_{k=1}^{N_{0}}$ uniformizying 
$((M;N_{0}),\mathcal{C})$ around the punctures. At the level of cohomology,
we can equivalently write 
\begin{equation}
\vartheta _{k}=\left[ \frac{\sqrt{\phi (k)}}{l(\partial (\rho ^{2}(k)))}-
\frac{\sqrt{\phi (N_{0})}}{l(\partial (\rho ^{2}(N_{0})))}\right] .
\end{equation}
The basis $\{\vartheta _{k}\}_{k=1}^{N_{0}-1}$ can be completed to $H^{1}(M,N_{0};\mathbb{C})$ 
by introducing also a basis $\{\varphi
(a)\}_{a=1}^{g}$ for $H^{1,0}(M)$, ($g$ being the genus of $M$) according to 
\begin{equation}
\int_{M}\varphi _{a}\wedge \overline{\varphi }_{b}=-i\delta _{ab}.
\label{hodmet}
\end{equation}
\ Note that 
\begin{equation}
\int_{M}\vartheta _{j}\wedge \overline{\varphi }_{a}=0,  \label{ortho}
\end{equation}
since $\alpha \in I^{1,1}$ iff $\int_{M}\alpha \wedge \overline{
\varphi }_{b}=0$. It follows that on $H^{1}(M,N_{0};\mathbb{R})$ we have
the (Hodge-Deligne) inner product 
\begin{equation}
(\alpha ,\beta )_{H-D}\doteq \sum_{p\in
\{p_{k}\}_{k=1}^{N_{0}}}RES_{p}(\alpha )RES_{p}(\beta )+\int_{M}\ast \alpha
\wedge \beta  \label{HDbilinear}
\end{equation}
for the (unique) harmonic representatives $\alpha $ and $\beta $ of elements 
$[\alpha ]$ and $[\beta ]$ in $H^{1}(M,N_{0};\mathbb{R})$, where $\ast $
denotes the Hodge dual and $RES_{p}(\cdot )$ is the residue map. By
extending such an inner product to a hermitian form on $H^{1}(M,N_{0};
\mathbb{C})$ one gets a (mixed Hodge) metric on $H^{1}(M,N_{0};\mathbb{C})$
\cite{zucker} \cite{pearlstein2}.
It is worth stressing that the bilinear form (\ref
{HDbilinear}) induces the non-degenerate bilinear forms 
\begin{equation}
(\alpha ,\beta )_{(H-D)}^{\parallel }\doteq \sum_{p\in
\{p_{k}\}_{k=1}^{N_{0}}}RES_{p}(\alpha )RES_{p}(\beta ),\;\text{on }\frac{
H^{1}(M,N_{0};\mathbb{R})}{H^{1}(M,\mathbb{R})}  \label{metridotta}
\end{equation}
and 
\begin{equation}
(\alpha ,\beta )_{(H-D)}^{\perp }\doteq \int_{M}\ast \alpha \wedge \beta ,\;
\text{on }H^{1}(M,\mathbb{R}).
\end{equation}

\bigskip

\subsection*{\normalsize{\bf 4.5. A regularized Dirichlet norm.}}
In a similar vein, let us consider the function $v$ defining the conformal class of the conical
geometry (\ref{cmetr}) around the generic puncture $p_{k}$, \emph{i.e.}, 
\begin{equation}
\left. v\right| _{U_{\rho ^{2}(k)}}=-\frac{\varepsilon (k)}{2\pi }\ln \left|
\zeta (k)\right| +u,  \label{confac}
\end{equation}
where the function $u$ is continuous and $C^{2}$ on $U_{\rho ^{2}(k)}-p_{k}$
and is such that, for $\zeta (k)\rightarrow 0$, $\left| \zeta (k)\right| 
\frac{\partial u}{\partial \zeta (k)}$, and $\left| \zeta (k)\right| \frac{
\partial u}{\partial \overline{\zeta (k)}}$ both $\rightarrow 0$, (see (\ref
{cmetr})). Since 
\begin{gather}
\ast dv|_{U_{\rho ^{2}(k)}}=-\sqrt{-1}\left( \frac{\partial v}{\partial
\zeta (k)}d\zeta (k)-\frac{\partial v}{\partial \overline{\zeta }(k)}d
\overline{\zeta }(k)\right) = \\
=-\pi \left( \frac{\varepsilon (k)}{2\pi }\right) \left( \frac{d\zeta (k)}{
2\pi \sqrt{-1}\zeta (k)}-\frac{d\overline{\zeta }(k)}{2\pi \sqrt{-1}
\overline{\zeta }(k)}\right) +\ast du,  \notag
\end{gather}
it follows that the polar part of \ $\ast dv$ is in $I^{1,1}$ (in general
the whole $\ast dv$ is not in $I^{1,1}$ since $\partial \overline{\partial }
u\neq 0$) and we can consider its Hodge-Deligne norm on $\frac{
H^{1}(M,N_{0};\mathbb{R})}{H^{1}(M,\mathbb{R})}$  
\begin{equation}
(\ast dv,\ast dv)_{H-D}^{\parallel }=\sum_{p\in
\{p_{k}\}_{k=1}^{N_{0}}}RES_{p}\left( \ast dv\right) RES_{p}\left( \ast
dv\right) =4\pi ^{2}\sum_{k=1}^{N_{0}}\left( \frac{\varepsilon (k)}{2\pi }
\right) ^{2},
\end{equation}
(the factor $4\pi ^{2}$ is inserted for later convenience). This latter
expression is directly related with a natural regularization of the
otherwise ill-defined Dirichlet's energy associated with $v$, (\emph{i.e.},
the $L^{2}$ norm of the gradient of $v$). Let $M_{\varrho }\doteq \cup
_{k=1}^{N_{0}}\Delta _{k}^{\varrho }$\ where $\Delta _{k}^{\varrho }\doteq
\left\{ \zeta (k)\in \mathbb{C}|\;\varrho \leq |\zeta (k)|\leq 1\right\} $,
(thus $M_{\varrho }$ is topologically $M$ with $N_{0}$ disks $c_{\varrho
}(k) $ of radius $\varrho $ around the punctures removed). Since $
\overline{\partial }(\frac{d\zeta (k)}{\zeta (k)})=0$ in $\Delta
_{k}^{\varrho }$, we have 
\begin{gather}
\frac{\sqrt{-1}}{2}\int_{M_{\varrho }}\frac{\partial v}{\partial \zeta }
\frac{\partial v}{\partial \overline{\zeta }}d\zeta \wedge d\overline{\zeta }
=\frac{1}{4}\int_{M_{\varrho }}dv\wedge \ast dv= \\
=\frac{1}{4}\int_{M_{\varrho }}du\wedge \ast du+\frac{1}{2}
\sum_{k=1}^{N_{0}}\left( -\frac{\varepsilon (k)}{2\pi }\right) \frac{\sqrt{-1
}}{2}\int_{\Delta _{k}^{\varrho }\cap M_{\varrho }}\left( \partial u\wedge 
\frac{d\overline{\zeta }(k)}{\overline{\zeta }(k)}-\overline{\partial }
u\wedge \frac{d\zeta (k)}{\zeta (k)}\right)  \notag \\
+\frac{1}{4}\sum_{k=1}^{N_{0}}\left( \frac{\varepsilon (k)}{2\pi }\right)
^{2}\frac{\sqrt{-1}}{2}\int_{\Delta _{k}^{\varrho }\cap M_{\varrho }}\frac{
d\zeta (k)\wedge \ d\overline{\zeta }(k)}{|\zeta (k)|^{2}}=  \notag \\
=\frac{1}{4}\int_{M_{\varrho }}du\wedge \ast du-\frac{1}{2}
\sum_{k=1}^{N_{0}}\left( \frac{\varepsilon (k)}{2\pi }\right) \frac{\sqrt{-1}
}{2}\oint_{\partial c_{\varrho }(k)}u\left( \frac{d\overline{\zeta }(k)}{
\overline{\zeta }(k)}-\frac{d\zeta (k)}{\zeta (k)}\right) -  \notag \\
-\frac{\pi }{2}\ln \varrho \sum_{k=1}^{N_{0}}\left( \frac{\varepsilon (k)}{
2\pi }\right) ^{2},  \notag
\end{gather}
where $\partial c_{\varrho }(k)$ is the circle (with positive orientation)
of radius $\varrho $ around the generic puncture $p_{k}$. Thus, we get 
\begin{gather}
\frac{\sqrt{-1}}{2}\int_{M_{\varrho }}\frac{\partial v}{\partial \zeta }
\frac{\partial v}{\partial \overline{\zeta }}d\zeta \wedge d\overline{\zeta }
+\frac{\pi }{2}\ln \varrho \sum_{k=1}^{N_{0}}\left( \frac{\varepsilon (k)}{
2\pi }\right) ^{2}= \\
\frac{\sqrt{-1}}{2}\int_{M_{\varrho }}\frac{\partial v}{\partial \zeta }
\frac{\partial v}{\partial \overline{\zeta }}d\zeta \wedge d\overline{\zeta }
+\frac{1}{8\pi }\ln \varrho \,(\ast dv,\ast dv)_{H-D}^{\parallel }=  \notag
\\
=\frac{1}{4}\int_{M_{\varrho }}du\wedge \ast du-\frac{1}{2}
\sum_{k=1}^{N_{0}}\left( \frac{\varepsilon (k)}{2\pi }\right) \frac{\sqrt{-1}
}{2}\oint_{\partial c_{\varrho }(k)}u\left( \frac{d\overline{\zeta }(k)}{
\overline{\zeta }(k)}-\frac{d\zeta (k)}{\zeta (k)}\right) .  \notag
\end{gather}
Since the right hand side of this expression is well-defined in the $\varrho
\rightarrow 0^{+}$ limit, we can define a regularized $L^{2}$-norm of $dv$
according to 
\begin{equation}
\int_{M}^{reg}dv\wedge \ast dv\doteq \lim_{\varrho \rightarrow 0^{+}}\left[
\int_{M_{\varrho }}dv\wedge \ast dv+\frac{1}{2\pi }\ln \varrho \,(\ast
dv,\ast dv)_{H-D}^{\parallel }\right] .
\end{equation}
We can explicitly rewrite \noindent $\int_{M}^{reg}dv\wedge \ast dv$ as the
integral over $M$ of a $(1,1)$ current by noticing that 
\begin{gather}
\frac{1}{2}\sum_{k=1}^{N_{0}}\left( \frac{\varepsilon (k)}{2\pi }\right) 
\frac{\sqrt{-1}}{2}\oint_{\partial c_{\varrho }(k)}u\left( \frac{d\overline{
\zeta }(k)}{\overline{\zeta }(k)}-\frac{d\zeta (k)}{\zeta (k)}\right) = \\
=-\frac{\pi }{2}\sum_{k=1}^{N_{0}}\left( \frac{\varepsilon (k)}{2\pi }
\right) \int_{c_{\varrho }(k)}u\left( \partial \left( \frac{d\overline{\zeta 
}(k)}{2\pi \sqrt{-1}\overline{\zeta }(k)}\right) -\overline{\partial }\left( 
\frac{d\zeta (k)}{2\pi \sqrt{-1}\zeta (k)}\right) \right) +  \notag \\
+\frac{1}{2}\sum_{k=1}^{N_{0}}\left( \frac{\varepsilon (k)}{2\pi }\right) 
\frac{\sqrt{-1}}{2}\int_{c_{\varrho }(k)}\partial u\wedge \left( \frac{d
\overline{\zeta }(k)}{\overline{\zeta }(k)}\right) -\overline{\partial }
u\wedge \left( \frac{d\zeta (k)}{\zeta (k)}\right) =  \notag \\
=\pi \sum_{k=1}^{N_{0}}\left( \frac{\varepsilon (k)}{2\pi }\right)
\int_{M}u\delta _{p_{k}}\frac{\sqrt{-1}}{2}d\zeta (k)\wedge d\overline{\zeta 
}(k)+  \notag \\
+\frac{1}{2}\sum_{k=1}^{N_{0}}\left( \frac{\varepsilon (k)}{2\pi }\right) 
\frac{\sqrt{-1}}{2}\int_{c_{\varrho }(k)}\partial u\wedge \left( \frac{d
\overline{\zeta }(k)}{\overline{\zeta }(k)}\right) -\overline{\partial }
u\wedge \left( \frac{d\zeta (k)}{\zeta (k)}\right) .  \notag
\end{gather}
Under the stated hypotheses on $u$, the integrals over the disks $
c_{\varrho }(k)$ go uniformly to 0 as $\varrho \rightarrow 0^{+}$, and we
can write 
\begin{equation}
\int_{M}^{reg}dv\wedge \ast dv=\int_{M}du\wedge \ast du+4\pi
\sum_{k=1}^{N_{0}}\left( -\frac{\varepsilon (k)}{2\pi }\right)
\int_{M}u\delta _{p_{k}}\frac{\sqrt{-1}}{2}d\zeta (k)\wedge d\overline{\zeta 
}(k).
\end{equation}
More generally, if $v_{1}$ and $v_{2}$ are two locally summable functions on 
$((M;N_{0}),\mathcal{C})$ with local behavior 
\begin{eqnarray}
\left. v_{1}\right| _{U_{\rho ^{2}(k)}} &=&-\frac{\varepsilon _{1}(k)}{2\pi }
\ln \left| \zeta (k)\right| +u_{1}, \\
\left. v_{2}\right| _{U_{\rho ^{2}(k)}} &=&-\frac{\varepsilon _{2}(k)}{2\pi }
\ln \left| \zeta (k)\right| +u_{2},  \notag
\end{eqnarray}
then we can define a regularized inner product of their gradients according
to 
\begin{gather}
\int_{M}^{reg}dv_{1}\wedge \ast dv_{2}\doteq \lim_{\varrho \rightarrow
0^{+}} \left[ \int_{M_{\varrho }}dv_{1}\wedge \ast dv_{2}+\frac{1}{2\pi }\ln
\varrho \,(\ast dv_{1},\ast dv_{2})_{H-D}^{\parallel }\right] =
\label{inner} \\
=\int_{M}du_{1}\wedge \ast du_{2}+2\pi \sum_{k=1}^{N_{0}}\int_{M}\left( -
\frac{\varepsilon _{1}(k)}{2\pi }u_{2}-\frac{\varepsilon _{2}(k)}{2\pi }
u_{1}\right) \delta _{p_{k}}\frac{\sqrt{-1}}{2}d\zeta (k)\wedge d\overline{
\zeta }(k).  \notag
\end{gather}

\subsection*{\normalsize{\bf 4.6. The regularized Liouville action.}} 
The conformal class of the metric representing the Regge triangulation
associated with $((M;N_{0}),\mathcal{C})$ is given by $
ds^{2}=e^{2v}ds_{0}^{2}$ where $ds_{0}^{2}$ is a smooth metric on $M$ and
where the conformal factor $v$ around the generic puncture $p_{k}$, is
provided by (\ref{confac}). The Gaussian curvature $K$ of $ds^{2}$ is
related to the Gaussian curvature $K_{0}$ of the smooth metric $ds_{0}^{2}$
by the relation 
\begin{equation}
KdA=K_{0}dA_{0}-d\ast dv,
\end{equation}
where $dA_{0}$ and $dA$ are the area elements of $ds_{0}^{2}$ and $ds^{2}$,
respectively. By interpreting $d\ast dv$ as a $(1,1)$ current, a formal
direct computation yields 
\begin{align}
d\ast dv& =\pi \sum_{k=1}^{N_{0}}\left( -\frac{\varepsilon (k)}{2\pi }
\right) \left[ \overline{\partial }\left( \frac{d\zeta (k)}{2\pi \sqrt{-1}
\zeta (k)}\right) -\partial \left( \frac{d\overline{\zeta }(k)}{2\pi \sqrt{-1
}\overline{\zeta }(k)}\right) \right] +2\sqrt{-1}\partial \overline{\partial 
}u, \\
& =-2\pi \sum_{k=1}^{N_{0}}\left( -\frac{\varepsilon (k)}{2\pi }\right) 
\frac{\sqrt{-1}}{2}\delta _{p_{k}}d\zeta (k)\wedge d\overline{\zeta }(k)+2
\sqrt{-1}\partial \overline{\partial }u.  \notag
\end{align}
From which we get the (inhomogeneous) Liouville equation associated with the
conformal metric $ds^{2}$ with conical singularities $\{-\frac{\varepsilon
(k)}{2\pi }\}$, \emph{i.e.} 
\begin{equation}
-4\partial _{\zeta }\partial \overline{_{\zeta }}u=e^{2v}K-K_{0}-2\pi
\sum_{k=1}^{N_{0}}\left( -\frac{\varepsilon (k)}{2\pi }\right) \delta
_{p_{k}}.  \label{liouville}
\end{equation}
Note that, upon integrating (\ref{liouville}) over $M$ we have 
\begin{equation}
\frac{1}{2\pi }\int_{M}e^{2v}KdA_{0}=\frac{1}{2\pi }\int_{M}K_{0}dA_{0}+
\sum_{k=1}^{N_{0}}\left( -\frac{\varepsilon (k)}{2\pi }\right) ,
\end{equation}
which, since $\frac{1}{2\pi }\int_{M}K_{0}dA_{0}=\chi (M)$ and $
\sum_{k=1}^{N_{0}}\left( -\frac{\varepsilon (k)}{2\pi }\right) =-\chi (M)$
-(see (\ref{rediv})), is a restatement of the vanishing
of the Euler class (\ref{Euclass}) of $((M;N_{0}),\mathcal{C})$.

From the definition of regularized inner product (\ref{inner}) it is easily
verified that formally (\ref{liouville}) can be obtained as the
Euler-Lagrange equation of the functional 
\begin{equation}
S(v|\{\varepsilon (h)\})\doteq \frac{1}{4}\int_{M}^{reg}dv\wedge \ast
dv-\int_{M}e^{2v}KdA_{0}+2\int_{M}vK_{0}dA_{0},
\end{equation}
by considering variations $v_{t}$ of the form $v_{t}=v+th$, where the
function $h$ is continuous and $C^{2}$ on each $U_{\rho ^{2}(k)}-p_{k}$ and
is such that, for $\zeta (k)\rightarrow 0$, $\left| \zeta (k)\right| \frac{
\partial h}{\partial \zeta (k)}$, and $\left| \zeta (k)\right| \frac{
\partial h}{\partial \overline{\zeta }(k)}$ both $\rightarrow 0$. The
functional $S(v|K,K_{0})$ is basically a compact rewriting of the Takhtajan
and Zograf regularized Liouville action \cite{zograf2}.

\bigskip 

\noindent{\bf Acknowledgements}

\vspace{0.1cm}

\noindent Mauro Carfora warmly thanks professor G. Ferrarese for the kind invitation to
a most interesting and enjoiable meeting. This work was supported in part by the Ministero dell'Universita' e della
Ricerca Scientifica under the PRIN project \emph{The geometry of integrable
systems.}

\end{document}